\documentclass[fleqn,usenatbib]{mnras}

\usepackage{newtxtext,newtxmath}

\usepackage[T1]{fontenc}
\usepackage{ae,aecompl}

\usepackage{graphicx} 
\usepackage{amsmath} 
\usepackage{amssymb} 
\usepackage{amsfonts}
\usepackage{amssymb}

\title[The effects of unresolved DDs in the WD LF]{The effects of unresolved double-degenerates in the white dwarf luminosity function}

\author[Rebassa-Mansergas et al.]{A. Rebassa-Mansergas$^{1,2}$\thanks{E-mail: alberto.rebassa@upc.edu}, S. Toonen$^{3}$, S. Torres$^{1,2}$ and P. Canals $^{1}$
\\
$^{1}$ Departament de F\'{\i}sica, Universitat Polit\`{e}cnica de Catalunya, c/Esteve Terrades 5, 08860 Castelldefels, Spain\\
$^{2}$ Institut d'Estudis Espacials de Catalunya, Ed. Nexus-201, c/Gran Capit\`a 2-4, 08034 Barcelona, Spain\\
$^{3}$ School of Physics and Astronomy, University of Birmingham, Edgbaston, Birmingham B15 2TT, United Kingdom
}

\date{Accepted XXX. Received YYY; in original form ZZZ}

\pubyear{2019}

\begin{document}
\label{firstpage}
\pagerange{\pageref{firstpage}--\pageref{lastpage}}
\maketitle

\begin{abstract}
We  perform an  analysis  of the  single white  dwarf  and the  double
degenerate binary  populations in the solar  neighbourhood following a
population synthesis approach to investigate the effects of unresolved
double degenerates in the white dwarf luminosity function. We consider
all  unresolved synthetic  binaries to  be associated  with fictitious
effective temperatures and surface gravities  that are obtained in the
same way as if these objects were observed as single point sources. We
evaluate the effects  of unresolved double white  dwarfs assuming that
the synthetic  samples are ``observed'' both  by the magnitude-limited
SDSS and the volume-limited \emph{Gaia} surveys, the latter limited to
a distance  of no more  than 100\,pc. We  find that, for  our standard
model, the impact of unresolved  double degenerates in the white dwarf
luminosity  function derived  from  the \emph{Gaia}  sample is  nearly
negligible. Unresolved  double degenerates are hence  expected to have
no effect on the  age of the Galactic disc, nor  on the star formation
history from this population. However, for the SDSS sample, the effect
of unresolved double  degenerates is significant at  the brighter bins
(M$_\mathrm{bol}$<6.5 mag), with the fraction of such systems reaching
$\simeq$40\% of the total white dwarf population at M$_\mathrm{bol}=$6
mag.  This  indicates unresolved double degenerates  may influence the
constraints on the star formation  history derived from the SDSS white
dwarf sample.
\end{abstract}




\section{Introduction}

It is  well established that  white dwarfs  (WDs) are the  remnants of
main   sequence  stars   of  masses   up  to   $\simeq8-11\,M_{\odot}$
\citep[e.g.][]{Siess2007}.  Thus, WDs  are by far the  most common end
product of stellar evolution. WDs are supported by the pressure of the
degenerate  electrons  in  their  interiors   and  are  subject  to  a
relatively well understood cooling process that drives their evolution
\citep[e.g.][]{Althaus2010a, Renedo2010, Tremblay2011, Camisassa2016}.
The theoretical  cooling sequences allow deriving  accurate WD cooling
ages from  the observationally  determined effective  temperatures and
surfaces  gravities.   By  making  use  of  an  initial-to-final  mass
relation  \citep[e.g.][]{Andrews2015,  Cummings2018}  and  adopting  a
metallicity value,  one can also  derive the WD progenitor  masses and
their  lifetimes  from  the  appropriate  main  sequence  evolutionary
sequences \citep[e.g.   BaSTI;][]{Pietrinferni2004}.  This  yields the
total WD ages by simply adding  the cooling times to the main sequence
progenitor  lifetimes.    Hence,  WDs  have  been   used  as  reliable
cosmochronometers  for  the  determination  of  ages.   Some  relevant
examples of  using WDs  as cosmochronometers include  constraining the
age-metallicity  relation of  the Galactic  disk \citep{Rebassa2016a},
the  age-rotation-activity relation  of low-mass  main sequence  stars
\citep{Rebassa2013,  Skinner2017}   or  the   age-velocity  dispersion
relation  of the  Galactic disk  \citep{Anguiano2017}.  It  has to  be
emphasised  however  that the  most  efficient  way  of using  WDs  as
reliable  age  indicators  is  via  the study  of  the  WD  luminosity
function.

The WD luminosity  function is defined as the number  of WDs per cubic
parsec     as    a     function     of     unit    luminosity     (see
\citealt{Garcia-Berro2016} for  a recent  review). In  particular, the
analysis of the WD luminosity  function has provided age estimates for
the different  components of the Galactic  disk \citep{wingetetal87-1,
  garcia-berroetal88-1,  fontaineetal01-1,   Torres2016,  Kilic2017a},
constrained the  local star formation  rate \citep{diaz-pintoetal94-1,
  rowell13-1,  Isern2019}  and provided  the  ages  of both  open  and
globular  clusters \cite[e.g.][]{GBerro2010,  Jeffery2011, Hansen2013,
  Torres2015}.  However,  it is  important to  note that  the observed
samples from which the WD  luminosity functions are constructed suffer
from several drawbacks, which requires addressing the following issues
\citep{Garcia-Berro2016}:

\begin{itemize}
    \item An ultra  large WD sample is needed, of  the order of 10$^5$
      stars,  to  map  the   luminosity  function  with  the  smallest
      luminosity bins as possible.
    \item  Precise   parallaxes  and  proper  motions   are  required,
      substantially  better than  1\,mas and  1\,mas/yr, respectively.
      This is necessary for the identification of WD components of the
      different Galactic disks and the halo.  For example, the precise
      parallaxes  and  proper  motion  measurements  obtained  by  the
      \emph{Gaia}  mission have  permitted to  classify the  different
      Galactic     components    in     a    volume-limited     sample
      \citep{Torres2019}.
    \item Spectroscopic  identifications of sufficient  resolution for
      velocity determinations are mandatory,  not only for cataloguing
      the stars as  members of the different components,  but also for
      kinematic    studies   that    identify   Galactic    structures
      \citep{Torres2019b}.
    \item Improved  atmospheric models for  very cool WDs  are desired
      for a proper determination of their stellar parameters.
    \item Better categorization and treatment of selection effects are
      needed  to   account  for  all  possible   observational  biases
      affecting the samples.
    \item Quantifying the effects of  unresolved binaries is needed to
      understand   to  what   extent   the   luminosity  function   is
      contaminated.
\end{itemize}

Fortunately,  the  first  two  issues   are  being  addressed  by  the
\emph{Gaia}  mission \citep{Gaia},  which  is providing  unprecedented
samples    of    WDs,    both   magnitude-    \citep[][over    250,000
  objects]{Gentile2019}   and    volume-limited   \citep{Hollands2018,
  Jimenez2018}, along  with very  accurate photometry  and astrometry.
Moreover, WD  radial velocities of uncertainties  $\simeq$10\,km/s can
be   determined   from   medium   resolution   spectroscopy   if   the
signal-to-noise  ratio   is  sufficiently  high  (see,   for  example,
\citealt{Anguiano2017}). We  are also  improving our  understanding of
atmospheric models  for cool  WDs \citep{Blouin2019,  Tremblay2019} as
well as WD evolutionary sequences \citep{Camisassa2016, Camisassa2018,
  Degeronimo2018}.   With these  improvements  it is  now possible  to
efficiently   characterise   most  observational   selection   effects
\citep[e.g.][]{Rebassa2015a,  Cojocaru17}.   However, the  effects  of
unresolved binaries  such as double  degenerates in the  WD luminosity
function have not been yet explored in full detail. This is the aim of
the presented work.

We  study the  effects  of  unresolved double  degenerates  in the  WD
luminosity function  by means of  a population synthesis  simulator of
the  single WD  and the  double  degenerate populations  in the  solar
neighbourhood. We consider the synthetic samples as if they would have
been  observed  by  the  magnitude-limited Sloan  Digital  Sky  Survey
\citep[SDSS;][]{Yorketal00,   Stoughtonetal02}  and   the  \emph{Gaia}
mission, which provides a volume-limited WD sample for distances up to
100 pc \citep{Jimenez2018}.

\section{The population synthesis code}
\label{s-code}

We  used the  population  synthesis  code \textit{SeBa}  \citep{Por96,
  Too12} to generate  synthetic populations of single  and double WDs.
In both cases we only considered hydrogen-rich (DA) atmospheres, which
are   by   far   the   most    common   spectral   types   among   WDs
\citep[][]{Kepler2019}.  We  followed the evolution of  the stars from
the  zero-age   main-sequence  to  the  current   epoch.   The  double
degenerate populations were those we used in \citet{Rebassa2019}.  For
a   full  description   of  the   model,  we   refer  the   reader  to
\citet{Rebassa2019}, whilst a full  review of the population synthesis
method used here can be found in \citet{Too14}.

For  the initial  population of  single stars,  we assumed  that their
masses followed the initial mass function of \cite{Kro93} with a range
between  0.1-100$M_{\odot}$. Based  on the  same population  synthesis
approach, 10-30\% of single WDs may form through binary system mergers
\citep{Too17,   Temmink2019}.   Other   studies  claim   fractions  of
$\simeq$10\% \citep{Maoz2018} and $\simeq$15\% \citep{Kilic2018}.  The
impact of such  mergers in the WD luminosity function  will be studied
in a  future publication.  Hence,  single WDs that result  from binary
system mergers were not taken into  account in this work.  The initial
mass function was also adopted for  deriving the initial masses of the
initially most massive star in the binary. The masses of the companion
stars were drawn from a uniform distribution in the mass ratio between
0 and 1 \citep{Rag10, Duc13, Ros14, Cojocaru17}. The lowest value used
for  the mass  ratio  was $1\times10^{-10}$.   The orbital  separation
followed  a uniform  logarithmic distribution  \citep{Abt83}, and  the
eccentricity  a  thermal  distribution \citep{Heg75}.   We  adopted  a
constant binary  fraction of 50\%  as is  appropriate for A-,  F-, and
G-type   stars   \citep{Rag10,   Duc13,  Moe17,   Ros14}   and   solar
metallicities.

The formation of double WDs was determined by the common-envelope (CE)
phase. This is a short phase in  the evolution of the system, in which
both stars share a single (that is common) envelope \citep[see ][for a
  comprehensive  review]{Iva13}. It  plays  an important  role in  the
formation of many compact binaries, as the orbit is expected to shrink
due to friction.  The CE-phase  is poorly understood, and therefore we
applied  two  different  models,   designated  by  $\alpha\alpha$  and
$\gamma\alpha$.   In the  former, the  modelling of  the CE-phase  was
based on the energy budget \citep{Pac76,Tut79,Web84,Liv88}:
\begin{equation}
\alpha \Delta E_{\rm orb} = E_{\rm gr} \equiv \frac{GMM_{\rm env}}{\lambda R},    
\end{equation}
where $\alpha$  is the  efficiency with  which orbital  energy $E_{\rm
  orb}$ is consumed to unbind the CE with binding energy $E_{\rm gr}$,
$M$  the mass  of the  donor star,  $M_{\rm env}$  the envelope  mass,
$\lambda$ the binding  energy parameter that depends  on the structure
of the primary star, and $R$ the  radius of the donor star. We adopted
$\alpha\lambda=2$ based  on the  reconstruction of  the last  phase of
mass  transfer  in  the  evolution  of  observed  double  WDs,  as  in
\citet{Nel00}.

In  our  other  model  ($\gamma\alpha$),  we  adopted  an  alternative
prescription  for the  CE-phase  if  the binary  did  not contain  any
compact object  and the system  was stable against  the Darwin-Riemann
instability   \citep{Dar1879}.   In   this   alternative  model,   the
angular-momentum budget was considered:
\begin{equation}
    \frac{\Delta J}{J_{\rm init}} = \gamma \frac{\Delta M}{M+m},
\end{equation}
where J  is the angular  momentum of the binary,  $m$ the mass  of the
companion, and $\gamma$ the efficiency  with which angular momentum is
used to  unbind the envelope.  The  model is also inspired  by the the
work  of   \citet{Nel00}.   Here  we  adopted   $\gamma=1.75$,  as  in
\citet{Nel01}.

We assumed  a star formation  history (SFH) appropriate for  the Milky
Way that  is dependent  on Galactrocentric  radius and  time following
\citet{Boi99}  (see  \citealt{Too13} for  a  full  description of  the
model).  The  absolute magnitudes  of the WDs  were obtained  from the
cooling sequences of pure hydrogen-rich atmospheres (i.e. DA WDs) from
\cite{Hol06},  \cite{Kow06} and  \cite{Tremblay2011}\footnote{See also
  http://www.astro.umontreal.ca/bergeron/CoolingModels.}. The absolute
magnitude  of a  WD was  converted into  apparent magnitude  using its
distance  (as  provided  by  the   SFH  model)  and  the  interstellar
extinction based on its location in the Galaxy \citep[see][]{Too13}.

\section{Unresolved double-degenerates}
\label{s-unresol}

The  population  synthesis  code  described in  the  previous  Section
provided two different WD populations, namely the single WD sample and
the   double-degenerate  population.    Among  the   double-degenerate
binaries, we were specially interested in unresolved pairs, since they
are   observed   as   ``single''    WDs.    The   stellar   parameters
observationally derived  from the combined fluxes  of these unresolved
double-degenerates  are not  real (they  could be  considered as  good
approximations for the  hotter and/or less massive WD if  it were much
more  luminous than  the  companion). Hence,  the  inclusion of  these
apparently single WDs contaminates  the luminosity function.  In order
to analyse the  level of contamination, it became  necessary to derive
the  luminosities  (or bolometric  magnitudes  in  our case)  for  the
unresolved double-degenerates  as they  would have been  obtained from
observations. We did this as follows.

The first step was to  evaluate which synthetic double-degenerates can
be considered as  unresolved.  To that end, we  calculated the angular
separation of each  binary using equation 12 of  \citet{Too17}.  As we
have already  mentioned, we focus our  study on the WD  populations as
they would have been observed by the SDSS and the \emph{Gaia} surveys.
We  thus considered  as unresolved  double-degenerates those  binaries
with  angular   separation  of  1"   or  less  for  the   SDSS  sample
\citep{Yorketal00}, and  of 0.5"  or less  for the  \emph{Gaia} sample
\citep{debruijne2015}. Once  the unresolved binaries  were identified,
we obtained  their synthetic  spectra (resulting  from summing  up the
individual   WD   fluxes)   following    the   method   described   in
\citet{Rebassa2019}.   In  summary,  the  spectra  of  the  individual
components  were obtained  interpolating  the corresponding  effective
temperature and  surface gravity  values on an  updated grid  of model
atmosphere  spectra  of  \citet{Koester10}, which  incorporates  model
spectra  down  to effective  temperatures  of  4000\,K. The  resulting
combined  synthetic spectra  were then  fitted with  the same  grid of
hydrogen-rich  WD model  atmosphere spectra  to derive  the fictitious
effective  temperatures  and   surface  gravities  \citep[for  details
  see][]{Rebassa2017}.  These values were  then interpolated in the WD
cooling sequences  of \citet{Renedo2010}  for deriving  the bolometric
magnitudes.  In cases where the  individual components were both below
5000\,K we could not trust the surface gravity values derived from the
fits  since no  Balmer  lines were  present in  the  spectra. Thus  we
derived  a  fictitious  M$_\mathrm{g}$  absolute  magnitude  from  the
combined  $g$ apparent  magnitude  of the  double  degenerate and  its
distance  (taking extinction  into account),  which together  with the
effective temperature (reliable from  the fit) provided the bolometric
magnitudes by interpolating in  the cooling sequences mentioned above.
We emphasise  that the bolometric  magnitude values derived  here were
obtained by  considering the  unresolved double-degenerates  as single
WDs and are therefore fictitious.

\section{The white dwarf luminosity function}
\label{s-lf}

We  followed  two  different  approaches to  evaluate  the  impact  of
unresolved double-degenerates in the WD luminosity function. First, we
considered   the    synthetic   populations    of   single    WD   and
double-degenerate binaries  as they  would have  been observed  by the
SDSS  magnitude-limited  survey, which  so  far  contains the  largest
number  of  spectroscopically  confirmed  WDs  \citep[for  the  latest
  updated catalogue  see][]{Kepler2019} from which  several luminosity
functions have been derived \citep{Hu2007, deGenaro2008, Rebassa2015}.
The second approach was to only take into account synthetic WDs nearer
than 100 pc,  since these are accessible by  the \emph{Gaia} satellite
and  can  be considered  to  be  members  of a  volume-limited  sample
\citep{Jimenez2018}.  In  both exercises we considered  the unresolved
synthetic WDs as single point sources of fictitious stellar parameters
and bolometric magnitudes (Section\,\ref{s-unresol}).

\subsection{The SDSS magnitude-limited sample}

The SDSS is  a magnitude-limited survey originally  aimed at obtaining
spectra of  quasars and galaxies  \citep{Yorketal00, Stoughtonetal02}.
The target selection algorithm was slightly modified during subsequent
data releases \citep{yannyetal09-1,  ahnetal12-1, alametal15-1}.  This
implies the spectroscopic catalogue of SDSS WDs is heavily affected by
selection  effects,   despite  the  fact   that  it  is   the  largest
spectroscopic WD sample to date. The  14th data release (DR14) of SDSS
compiles over 35,000 spectroscopically-identified WDs, the majority of
which  are  of  spectral  type  DA  (hydrogen-rich  atmospheres)  with
effective temperatures between  $\simeq$6000\,K and $\simeq$100,000\,K
\citep{Kepler2019}.

\begin{figure}
\includegraphics[angle=-90,width=\columnwidth]{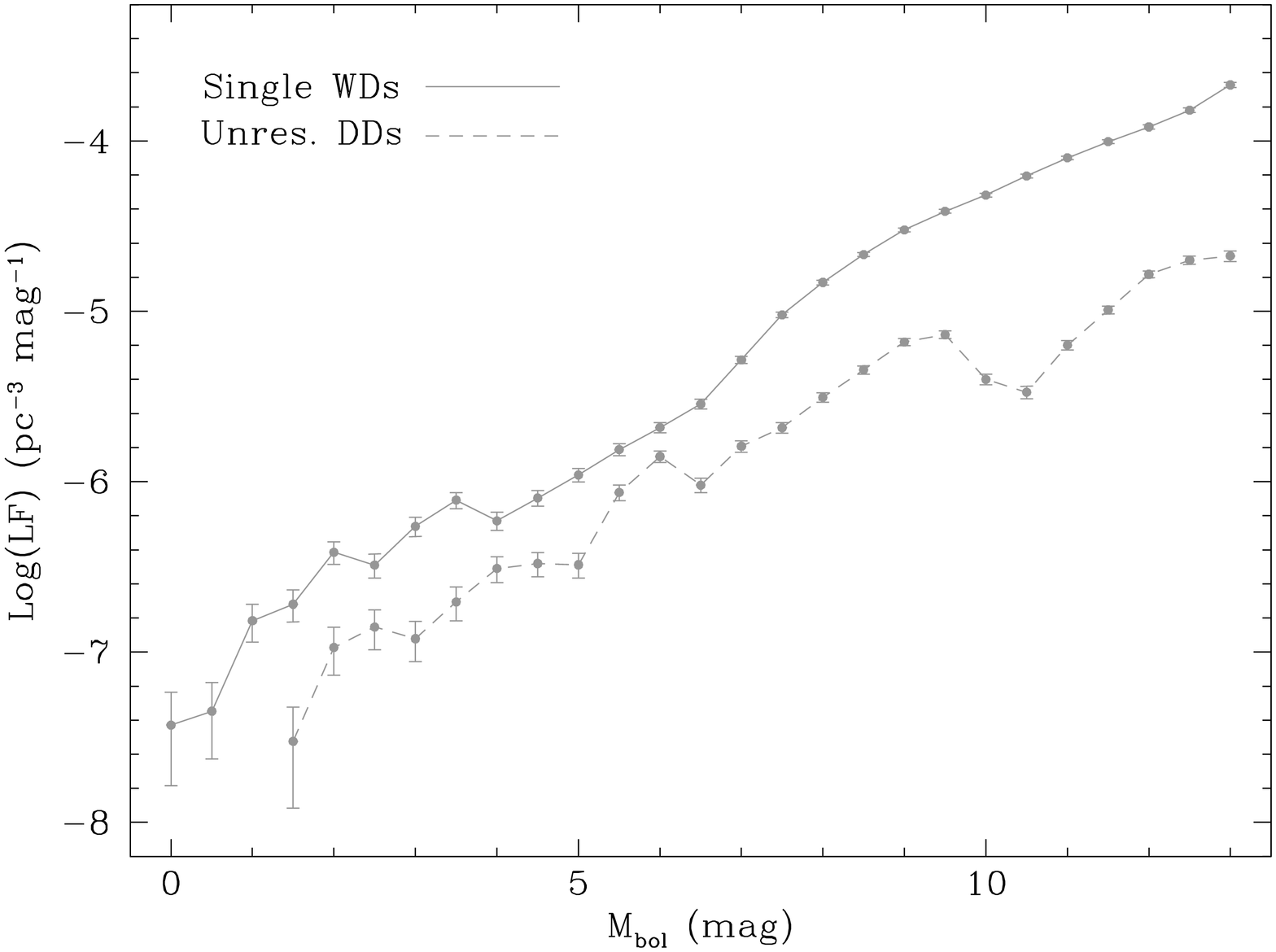}
\includegraphics[angle=-90,width=\columnwidth]{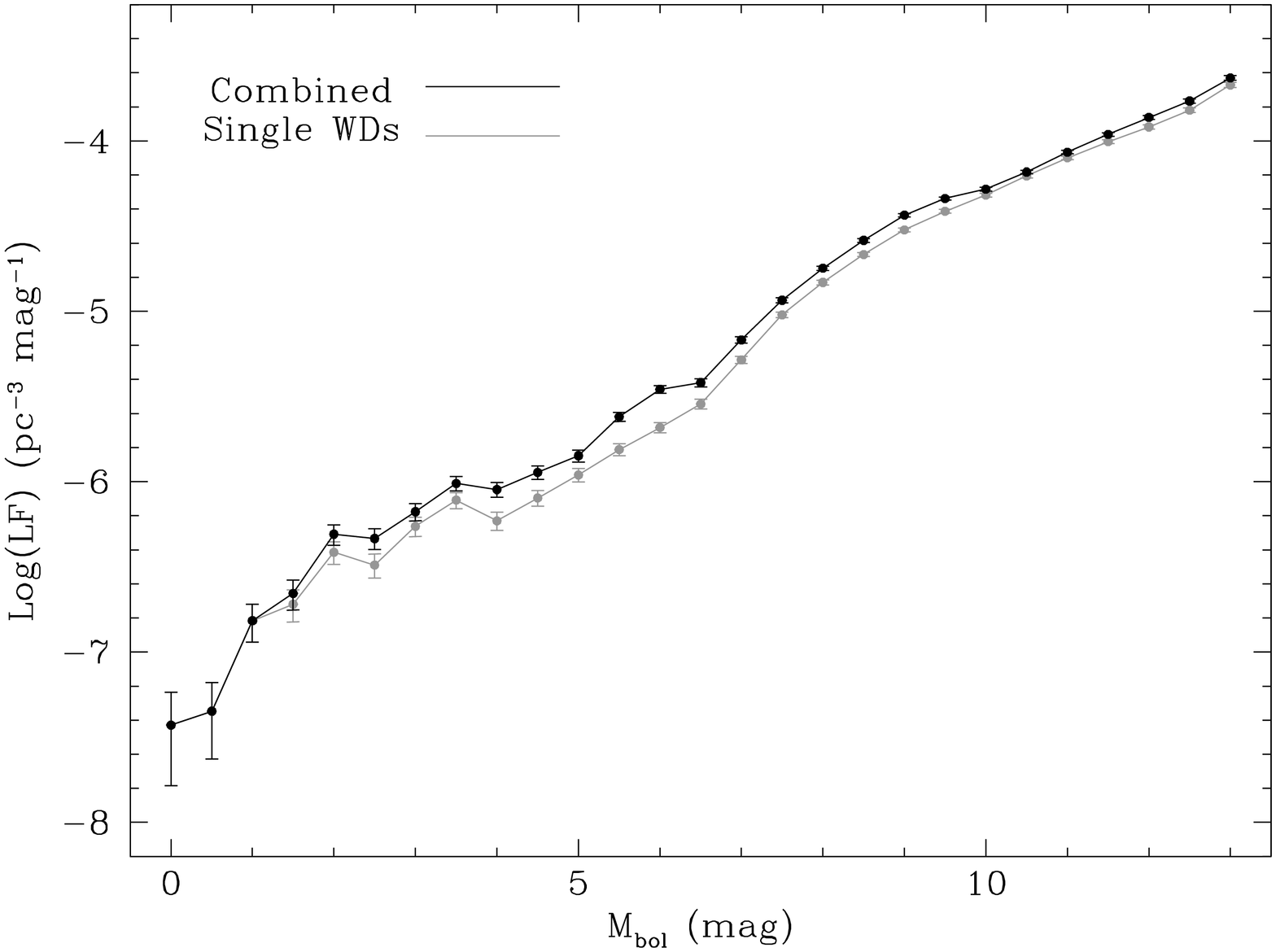}
\includegraphics[angle=-90,width=\columnwidth]{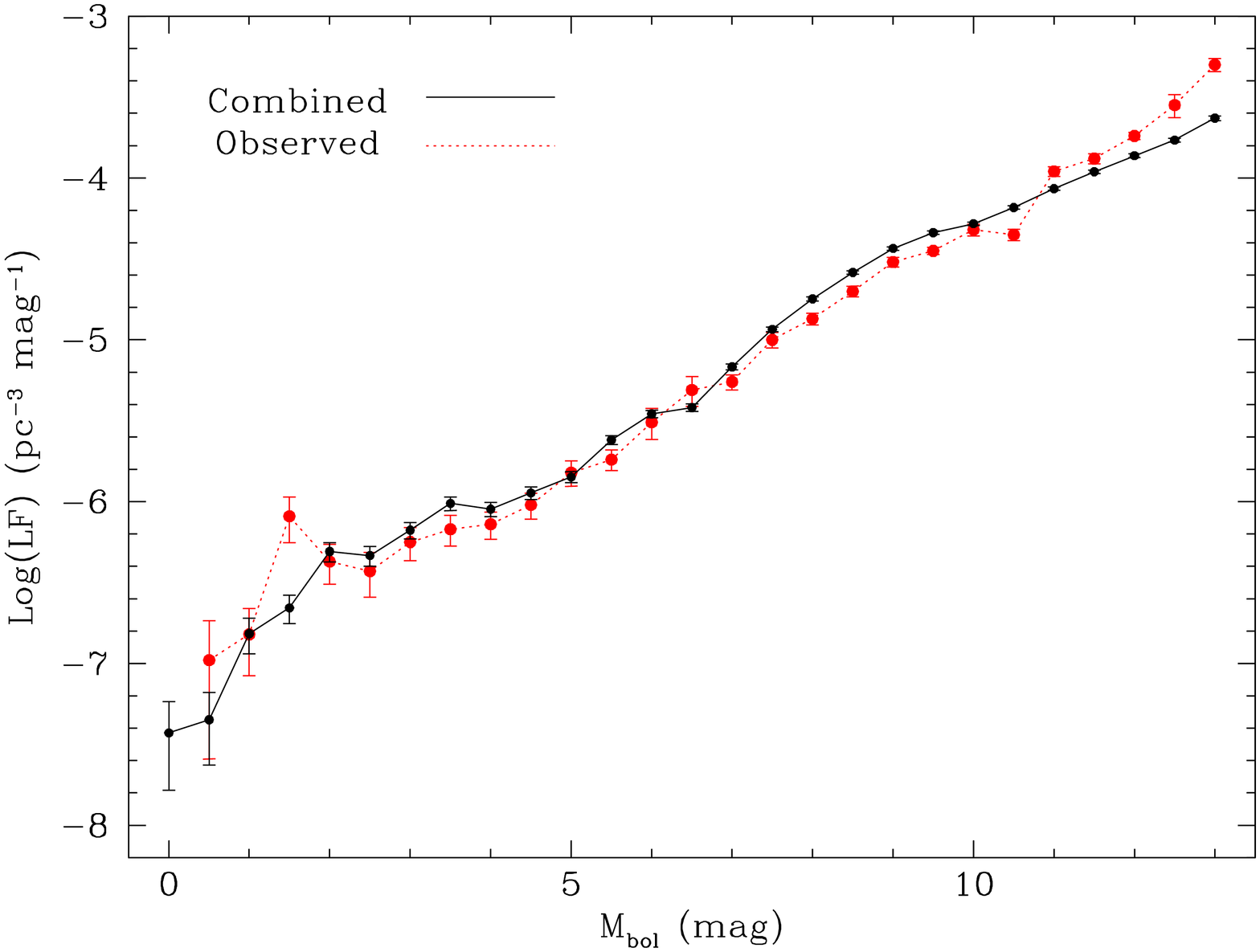}
    \caption{Top panel:  the synthetic  SDSS luminosity  functions for
      the  single  WD (solid  grey  line)  and the  unresolved  double
      degenerate (dashed grey line) samples obtained from our standard
      model.  Middle  panel: the combined luminosity  function (single
      WD  plus unresolved  double  degenerates; black  solid line)  as
      compared to  the one derived  from the single synthetic  SDSS WD
      sample  (grey  solid line)  when  adopting  our standard  model.
      Bottom  panel:   the  synthetic  combined   luminosity  function
      resulting from  assuming our  standard model (black  solid line)
      and  the observed  SDSS DA  WD luminosity  function (red  dotted
      line; \citealt{Rebassa2015}).}
\label{f-lfsdss1}
\end{figure}

Here we analyse our synthetic single  and binary WD samples as if they
were observed  by the SDSS survey  with the ultimate goal  of deriving
the   corresponding    luminosity   functions.    As    mentioned   in
Section\,\ref{s-unresol},  for the  binary samples  we calculated  the
angular separation of each source and filtered out all systems with an
angle larger  than 1", i.e. resolved  binaries by the SDSS  survey. We
calculated  the  magnitudes  of   the  unresolved  binaries  from  the
individual magnitudes  of each  binary component,  taking interstellar
extinction into account. Moreover,  we only considered those synthetic
WDs that fall  within the DR14 spectroscopic plates  (assuming a field
of   view    of   7\,deg$^2$   for   each    plate)   and   satisfying
14$\leq\,g\,\leq$19.1\,mag. $g=14$\,mag  is the lower  magnitude limit
of SDSS.  The  upper limit $g=19.1$\,mag was set due  to the fact that
fainter objects are generally associated  with noisy SDSS spectra that
were not used in the observed analysis of the SDSS luminosity function
\citep[see, for example,][]{Rebassa2015}. Moreover, we excluded WDs of
bolometric  magnitudes  fainter  than M$_\mathrm{bol}=$13  mag,  since
these objects  are not present in  the observed SDSS DA  WD luminosity
function \citep{Rebassa2015}. Finally, we applied the photometric cuts
in   the  SDSS   filters  for   spectroscopic  DA   WDs  provided   by
\citet{Girven2011}.   Thus, we  obtained  final  synthetic samples  of
20,842 single  WDs, 5,503  unresolved double degenerates  that evolved
adopting the  $\gamma\alpha$ formalism  for common envelope  and 4,674
unresolved  WD  binaries  that evolved  following  the  $\alpha\alpha$
formalism.

In order to  account for the selection effects  of a magnitude-limited
sample, we applied  a volume correction following  the 1/$V_{\rm max}$
method  \citep{schmidt68-1, green80-1}.   This  method calculates  the
maximum volume $V_\mathrm{WD}$  in which each synthetic  WD would have
been detected within the magnitude limits of the SDSS survey:

\begin{eqnarray}
\label{eq-vol}
V_\mathrm{WD} = V_\mathrm{max}-V_\mathrm{min} = {\sum_{i=1}^{n_{\rm plate}}}\,\frac{\omega_i}{4\pi}\int_{d_\mathrm{min}}^{d_\mathrm{max}}e^{-z/z_0}~4{\pi}r^2dr = \nonumber\\
= -\sum_{i=1}^{n_{\rm plate}} \frac{z_0 \times \omega_i}{\left |\sin{b}\right |}\left [ \left (r^2+\frac{2z_0}{\left |\sin{b}\right |}r+\frac{2z_0^2}{\left |\sin{b}\right |^2}  \right )\! e^{-\frac{r\left |\sin{b}\right |}{z_0}} \right ]_{d_\mathrm{min}}^{d_\mathrm{max}}
\end{eqnarray}

\noindent where  $b$ the  Galactic latitude of  the synthetic  WDs and
$\omega_i$ the solid  angle (in steradians) covered by  each SDSS DR14
plate.   The  quantity $e^{-z/z_0}$  is  essential to  consider  the
non-uniform distribution  of stars  in the direction  perpendicular to
the Galactic disc,  where $z=r \times \sin(b)$ the  WD's distance from
the   Galactic    plane   and    $z_0$=200\,pc   the    scale   height
\citep{Torres2019}. The lower/upper $g$  magnitude limits of each SDSS
plate  define  the   minimum/maximum  distance,  d$_\mathrm{min}$  and
d$_\mathrm{max}$, at which an object  belongs to the survey, hence the
minimum/maximum   volumes,  $V_\mathrm{min}$/$V_\mathrm{max}$,   where
$V_\mathrm{WD} = V_\mathrm{max}-V_\mathrm{min}$. The limits correspond
to the minimum/maximum $g$  magnitudes among all spectroscopic sources
observed by each plate \citep{Rebassa2015}.  The upper limits were set
to a maximum  of 19.1\,mag due to  the fact that we  restricted our WD
synthetic  samples to  objects of  $g$ magnitudes  brighter than  this
value.

As  previously  mentioned,  the  target selection  algorithm  of  SDSS
favoured  the detection  of hot  WDs of  similar colours  to those  of
quasars.  Therefore, the probability that  a spectrum was obtained for
a given  WD and contributes to  the WD luminosity function  depends on
its colour.  \citet{Rebassa2015}  applied a spectroscopic completeness
correction  to the  SDSS DA  WD observed  sample to  account for  this
observational  bias.   This  implied  the  observed  sample  could  be
considered  as  spectroscopically  complete.  Therefore,  rather  than
computing a  detection probability for  each of our synthetic  WDs and
applying the  corresponding spectroscopic completeness  correction, we
assumed that all synthetic WDs falling within the field of view of the
SDSS  DR14 plates  had an  associated spectrum.   In other  words, our
synthetic samples were also spectroscopically complete.

The synthetic single and double WD luminosity functions obtained using
our    adopted    model   (Section\,\ref{s-code}),    hereafter    our
\emph{standard   model},  are   illustrated  in   the  top   panel  of
Figure\,\ref{f-lfsdss1}.   We show  the luminosity  function from  the
$\gamma\alpha$ unresolved  double WD  population since  this formalism
best reproduces the observational data \citep{Nelemans2001, Too12}. In
both    cases,   the    uncertainties    are   calculated    following
\citet{boyle89-1}.   The  functions  display the  expected  monotonous
increase in the number of systems with fainter bolometric magnitudes.

\begin{figure}
\includegraphics[angle=-90,width=\columnwidth]{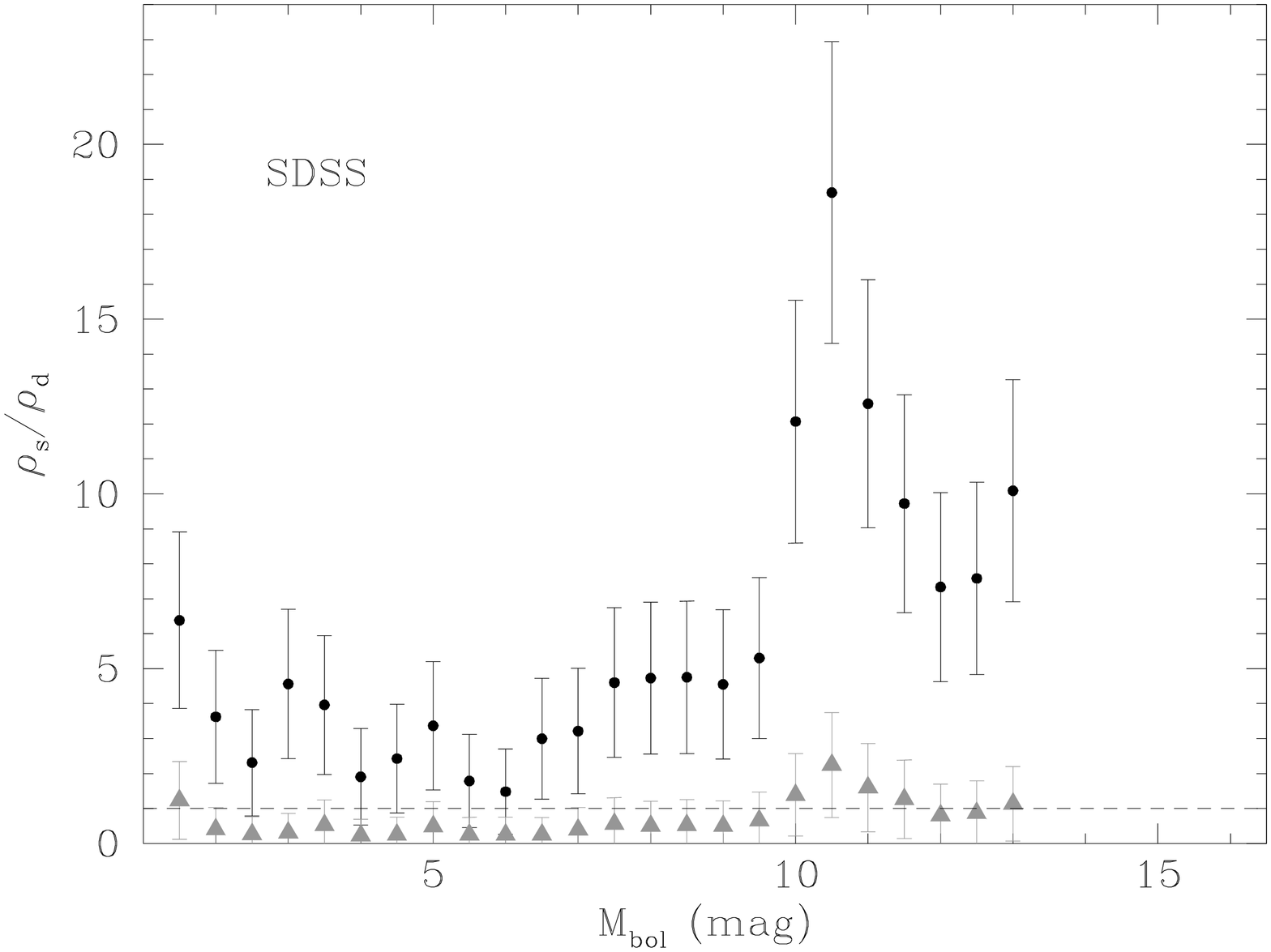}
\includegraphics[angle=-90,width=\columnwidth]{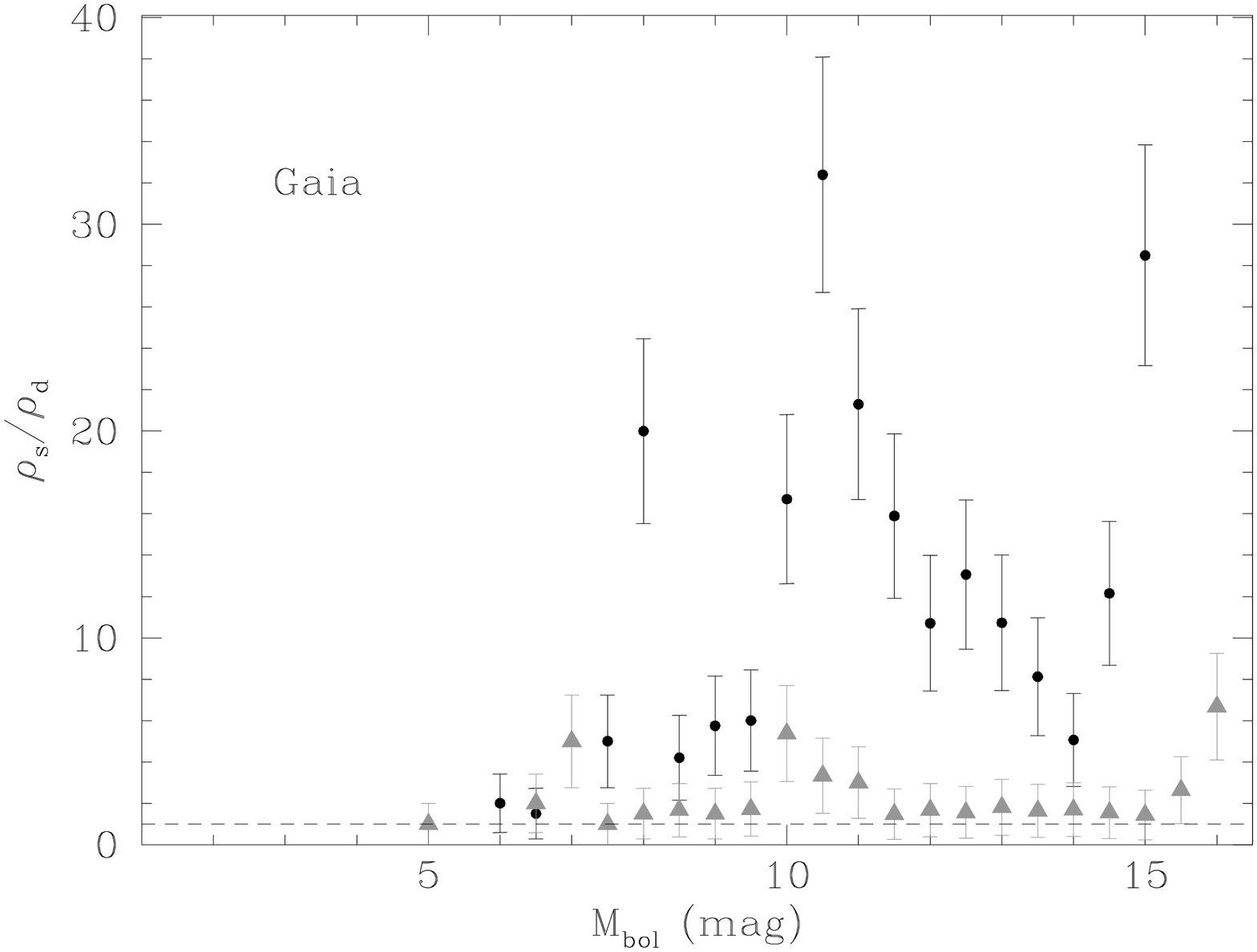}
    \caption{Top panel:  the fraction  between the space  densities of
      single WDs  and unresolved  double degenerates in  the synthetic
      SDSS sample  for our  standard model (black)  and for  the model
      producing the highest fraction  of unresolved double degenerates
      (gray; see Section\,\ref{s-morebin}). Bottom panel: the same but
      for  the \emph{Gaia}  synthetic samples.  The horizontal  dashed
      lines indicate fractions equal to 1.}
\label{f-fraction}
\end{figure}

In the middle panel of  Figure\,\ref{f-lfsdss1} we show the luminosity
function resulting from combining the single plus unresolved double WD
samples, which is compared to  the single WD luminosity function.  The
contribution  of  unresolved double  WDs  is  more pronounced  at  the
brighter bins,  especially between $\simeq$4 mag  < M$_\mathrm{bol}$ <
$\simeq$6 mag.   Previous studies  suggest that  low-mass, helium-core
WDs  that are  presumably  in  binary systems  are  the  cause for  an
increase in  the WD  luminosity function below  M$_\mathrm{bol}=$4 mag
\citep{Krzesinski2009, Torres2014}, which seems to be in line with our
results.   The other  bins  of the  luminosity  function are  scarcely
contaminated by the presence of unresolved double degenerate binaries.
This can  also be seen  in the top panel  of Figure\,\ref{f-fraction},
where we  show the ratio  of the space density  of single WDs  and the
space density of unresolved  double-degenerates.  The minimum fraction
value ($\simeq$1.5)  is achieved at M$_\mathrm{bol}=6$  mag.  In other
words,  unresolved   double  degenerates   can  represent,   at  most,
$\simeq$40\% of the objects with  M$_\mathrm{bol}$ around 6 mag. It is
also interesting to note that the ratio of the space density increases
considerably at M$_\mathrm{bol} \simeq 10.5$ mag, thus implying a drop
in the  space density  of unresolved double  degenerates.  We  need to
recall here that the derived M$_\mathrm{bol}$ values for such binaries
are  fictitious  and   are  not  directly  linked   to  real  physical
quantities.   Moreover, unresolved  system are  short period  binaries
whose evolution followed at least two episodes of mass transfer.  This
implies  that the  mass  distributions and  formation  times of  these
objects,  and  hence  their M$_\mathrm{bol}$  distributions,  are  not
expected to  follow the distributions  of both  single WDs and  WDs in
resolved binaries.

The  observed  SDSS luminosity  function  derived  by \citet[][;  this
  sample includes  5,857 apparently single  DA WDs from the  SDSS DR10
  catalogue  with $g\leq$19\,mag  and with  spectroscopic completeness
  correction  values  available]{Rebassa2015}  is illustrated  in  the
bottom  panel  of  Figure\,\ref{f-lfsdss1}  (red  dotted  line).   The
increase in  the observed space  density at M$_\mathrm{bol}=2$  mag is
very likely a  consequence of having very few WDs  at those bins, i.e.
we suffer  from low-number statistics. The  cut at M$_\mathrm{bol}=13$
mag is due to the fact that the lower effective temperature value that
can be determined for a  spectroscopic SDSS DA WDs is $\simeq$6000\,K.
Superimposed to the  observed SDSS luminosity function  we display the
one  obtained  from  combining  our  single  WD  plus  the  unresolved
double-degenerate synthetic samples.  The  combined function is a good
match   to   the   observed   one   except   at   the   fainter   bins
(M$_\mathrm{bol}=12,13$   mag)   where   our   simulations   seem   to
under-predict  the WD  space density.   Despite this  discrepancy, our
model assumptions seem to perform a good job at reproducing the the WD
and  WD binary  populations  in the  solar neighbourhood.   Therefore,
under  the assumptions  of  our standard  model,  the contribution  of
unresolved  double degenerates  appears to  be important  only at  the
brighter bins.  The extra contribution of apparently single WDs in the
luminosity  function  in  these  bins  (M$_\mathrm{bol}<6.5$  mag)  is
expected to  affect the derived  star formation history  from observed
SDSS WD samples \citep[e.g.][]{Hu2007}.   A typical 0.6 M$_{\odot}$ WD
with M$_\mathrm{bol}=6$ mag (where  we have the highest contamination)
has a total age of around 1.5 Gyr \citep{Camisassa2016}.  This implies
that a 40\%  contamination (we take the maximum possibe  value) in the
WD luminosity function for  M$_\mathrm{bol}<6.5$ mag translates into a
40\% uncertainty in the derived SFR for ages younger that 1.5 Gyr.

The  luminosity functions  obtained when  adopting the  $\alpha\alpha$
unresolved  double degenerate  sample were  nearly identical  to those
shown in Figure\,\ref{f-lfsdss1} for the $\gamma\alpha$ samples.  This
indicates  the common  envelope formalism  adopted plays  no important
role  in shaping  the luminosity  function.  Note  that modifying  the
particular values of the CE efficiency $\alpha$ for the $\alpha\alpha$
formalism  and   the  $\gamma$   efficiency  for   the  $\gamma\alpha$
prescription  do change  the fraction  of resulting  unresolved double
degenerates\footnote{The  fraction of  double WDs  amongst single  WDs
  differs by  a factor of  about 2  between the two  different models,
  with  the  $\gamma\alpha$ formalism  proving  the  larger number  of
  double  WDs  \citep{Rebassa2019}.    Unfortunately,  the  difference
  between  the  observationally  derived fractions  is  larger,  which
  implies  we  cannot use  them  yet  to constrain  binary  evolution.
  \citet{Too17} (using the same models) discuss the prospects for this
  with  the upcoming  large  and homogeneously  selected samples  from
  \emph{Gaia}.}.  However,  the assumed  values in our  standard model
are  already providing  the  highest possible  fractions, so  altering
these values  will only result in  a lower fraction of  double WDs and
hence an even lower impact in the WD luminosity function.

\subsection{The \emph{Gaia} 100\,pc volume-limited sample}

Thanks to the data provided by  the \emph{Gaia} mission we now have at
hand  the largest  sample of  WDs ever  obtained, which  includes over
250\,000 objects \citep{Gentile2019}. \citet{Jimenez2018} demonstrated
that the \emph{Gaia}  WD sample can be considered  as a volume-limited
one for  distances up to  100\,pc from the  Sun. This implies  one can
directly  obtain  the  WD  luminosity function  without  the  need  of
applying a  volume correction, i.e.  without the need of  applying the
1/$V_{\rm  max}$  method.   This   of  course  requires  deriving  the
bolometric  luminosities  of the  WDs,  which  currently can  only  be
obtained photometrically  since most  of the  \emph{Gaia} WDs  have no
spectra yet.  However, this exercise is subject to large uncertainties
due to  the fact that we  do not know  the spectral types of  the WDs,
hence  the need  of  adopting,  as a  first  approximation, those  for
hydrogen-rich WDs.

In  this  section   we  evaluate  the  impact   of  unresolved  double
degenerates  in  the  \emph{Gaia}  WD luminosity  function  that  will
eventually   be  obtained   once   all  \emph{Gaia}   WDs  have   been
spectroscopically observed.

\begin{figure}
\includegraphics[angle=-90,width=\columnwidth]{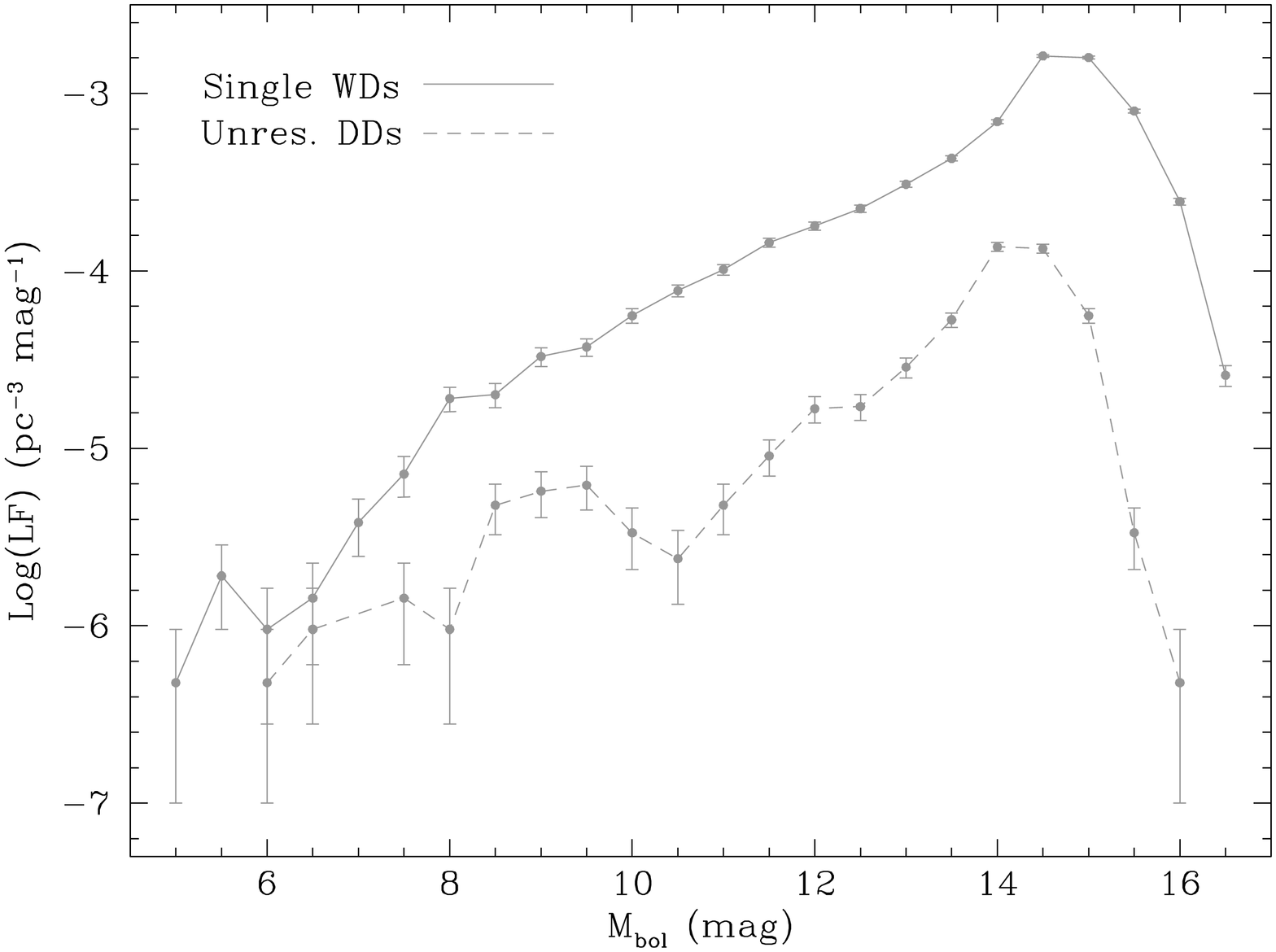}
\includegraphics[angle=-90,width=\columnwidth]{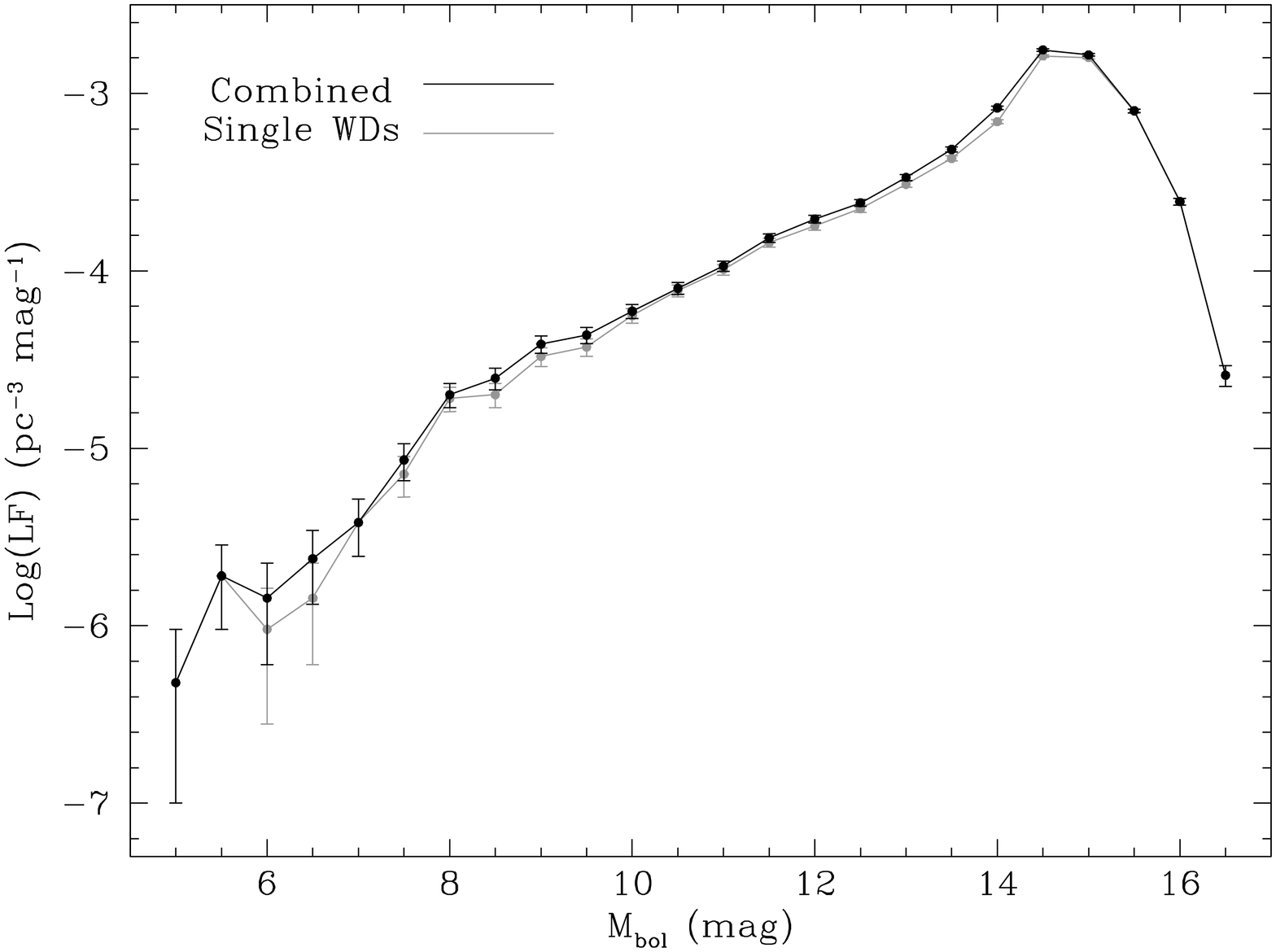}
\includegraphics[angle=-90,width=\columnwidth]{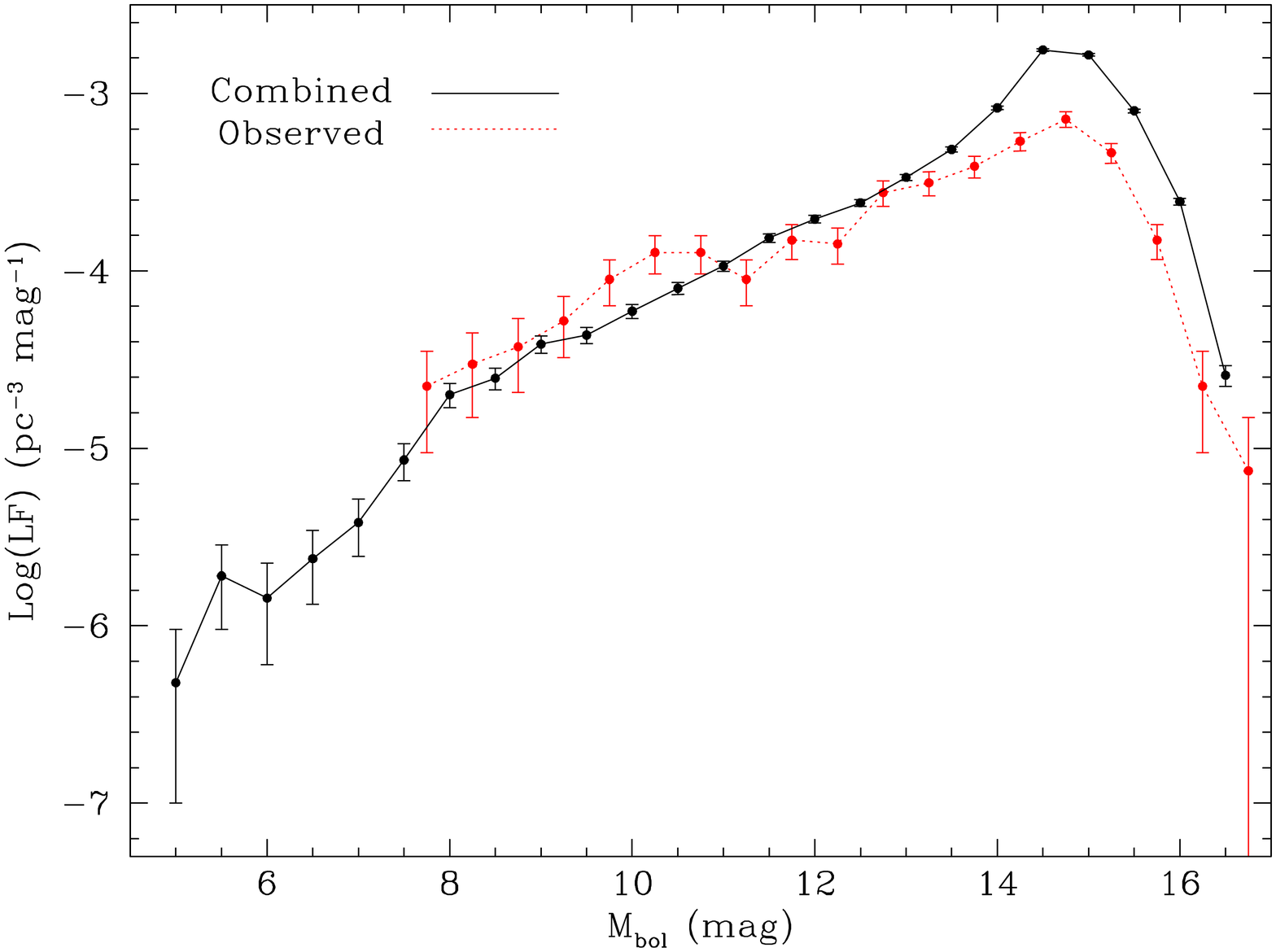}
    \caption{The   same  as   Figure\,\ref{f-lfsdss1}   but  for   the
      \emph{Gaia} samples.   The red dotted  line in the  bottom panel
      corresponds   to    the   observed   luminosity    function   of
      \citet{Limoges2015} for DA WDs within 40 pc of the Sun.}
\label{f-lfgaia1}
\end{figure}

To derive the corresponding  luminosity functions we considered single
WDs  and  unresolved  double  degenerates  of  effective  temperatures
(fictitious values  for the  latter sample) down  to 4000\,K  and with
distances within 100\,pc. The  chosen lower effective tempeature limit
was not  of much concern  since the Galaxy is  not old enough  to have
generated  many WDs  cooler than  this.  Indeed,  cooler WDs  are very
seldom  observed in  current volume-limited  samples \citep[e.g.   the
  40\,pc  sample from][]{Limoges2015}.   As  previously mentioned,  we
considered as unresolved binaries  those with angular separations less
than  0.5".  In  this case,  the number  of synthetic  single WDs  was
13,865 and the number of  unresolved double WDs for the $\gamma\alpha$
and $\alpha\alpha$ samples were 1017 and 670, respectively.

The   synthetic   functions   are   shown  in   the   top   panel   of
Figure\,\ref{f-lfgaia1},  where  the  errors are  calculated  assuming
Poisson statistics.   For the same reason  as for the SDSS  sample, we
only display the  $\gamma\alpha$ double WD binary  population. In both
cases, the  number of WDs  increases for fainter  bolometric magnitude
bins until  we reach M$_\mathrm{bol}\simeq$15\,mag. From  then on, the
space densities decrease  sharply, as expected from the  finite age of
the Galactic disk in a volume-limited  survey. It is worth noting that
the very bright M$_\mathrm{bol}$ bins ($<$5 mag) are absent of WDs due
to the  relatively small volume  sample we are  considering (100\,pc),
i.e. the larger  the volume the more  likely it is to  detect very hot
WDs.  The ratio of the space densities vs. M$_\mathrm{bol}$ for single
WDs and  unresolved double  degenerates is  illustrated in  the bottom
panel of Figure\,\ref{f-fraction}.  It becomes obvious that the number
of unresolved  double WDs  is much smaller  in the  \emph{Gaia} sample
than in  the SDSS sample,  a simple consequence  of the fact  that the
angular separation  of the  binaries that  the \emph{Gaia}  survey can
resolve is  half the  SDSS one.  Therefore,  the impact  of unresolved
double  degenerates in  the single  WD luminosity  function is  nearly
negligible (see  middle panel  of Figure\,\ref{f-lfgaia1}).   The only
region  where  there are  slight  differences  between the  luminosity
functions   for  single   WDs  and   the  combined   sample  is   near
M$_\mathrm{bol}=6$ mag (similar to the  result obtained using the SDSS
sample; Figure\,\ref{f-lfsdss1}).  Even here, the discrepancies in the
\emph{Gaia} case are within the  expected errors and can therefore not
be considered as  significant. Also worth noting is that  the ratio of
the space densities  for single WDs and  unresolved double degenerates
increases  considerably at  M$_\mathrm{bol} \simeq  10.5$ mag  (bottom
panel  of Figure\,\ref{f-fraction}).   This increase  is related  to a
drop in the space density  of unresolved double degenerates, an effect
that   was  also   observed  in   our  SDSS   sample  (top   panel  of
Figure\,\ref{f-fraction}).    The  fact   that  both   the  SDSS   and
\emph{Gaia} samples display the same  effect rules out the possibility
that the drop in space density  of unresolved double WDs at those bins
is a consequence of selection effects, since the \emph{Gaia} sample is
volume-limited.

For completeness, the bottom panel of Figure\,\ref{f-lfgaia1} displays
the synthetic  combined luminosity function and  the observed function
from \citet{Limoges2015}  for WDs within 40  pc of the Sun  (from this
sample  we only  considered  DA  WDs that  are  not  part of  resolved
binaries).   The agreement  between  the two  functions is  reasonable
considering  the observational  errors except  for M$_\mathrm{bol}$>14
mag, where  the synthetic function clearly  over-predicts the observed
one. This could be related to the fact that our standard model assumes
solar metallicities  rather than a dispersion  in metallicities around
the solar value \citep{Tononi2019}.

In  the same  way as  for the  SDSS sample,  the luminosity  functions
obtained by  adopting the $\alpha\alpha$ unresolved  double degenerate
sample were nearly identical to  those derived from the $\gamma\alpha$
samples.

We conclude  that, under  the assumptions of  our standard  model, the
star formation  history and  the age of  the Galactic  components that
will be  eventually derived from  the \emph{Gaia} 100\,pc  sample will
not be  affected by the  presence of unresolved double  degenerates in
the WD luminosity function.

\section{Discussion}

We have  adopted a  standard model based  on several  assumptions that
best  agree with  observational data  to reproduce  the population  of
unresolved double degenerates  (Section\,\ref{s-code}) and to evaluate
their      impact     in      the      WD     luminosity      function
(Section\,\ref{s-lf}). However,  the resulting fraction  of unresolved
double     WDs     clearly     depends    on     these     assumptions
\citep[e.g.][]{Claeys2014}.  Therefore,  there may exist  other models
that  yield   a  higher  number  of   unresolved  double  degenerates.
Moreover, resolved double WDs have so  far not been taken into account
in our analysis  despite the fact that,  observationally speaking, the
binary components  would be  considered as single  WDs and  hence also
contribute to some  extent to the WD luminosity  function. Finally, we
have  also  neglected the  contribution  from  WDs  that are  part  of
binaries with  main sequence companions.   In this section  we discuss
the impact of these issues.

\subsection{The fraction of unresolved double degenerates}
\label{s-morebin}

\begin{figure}
\includegraphics[angle=-90,width=\columnwidth]{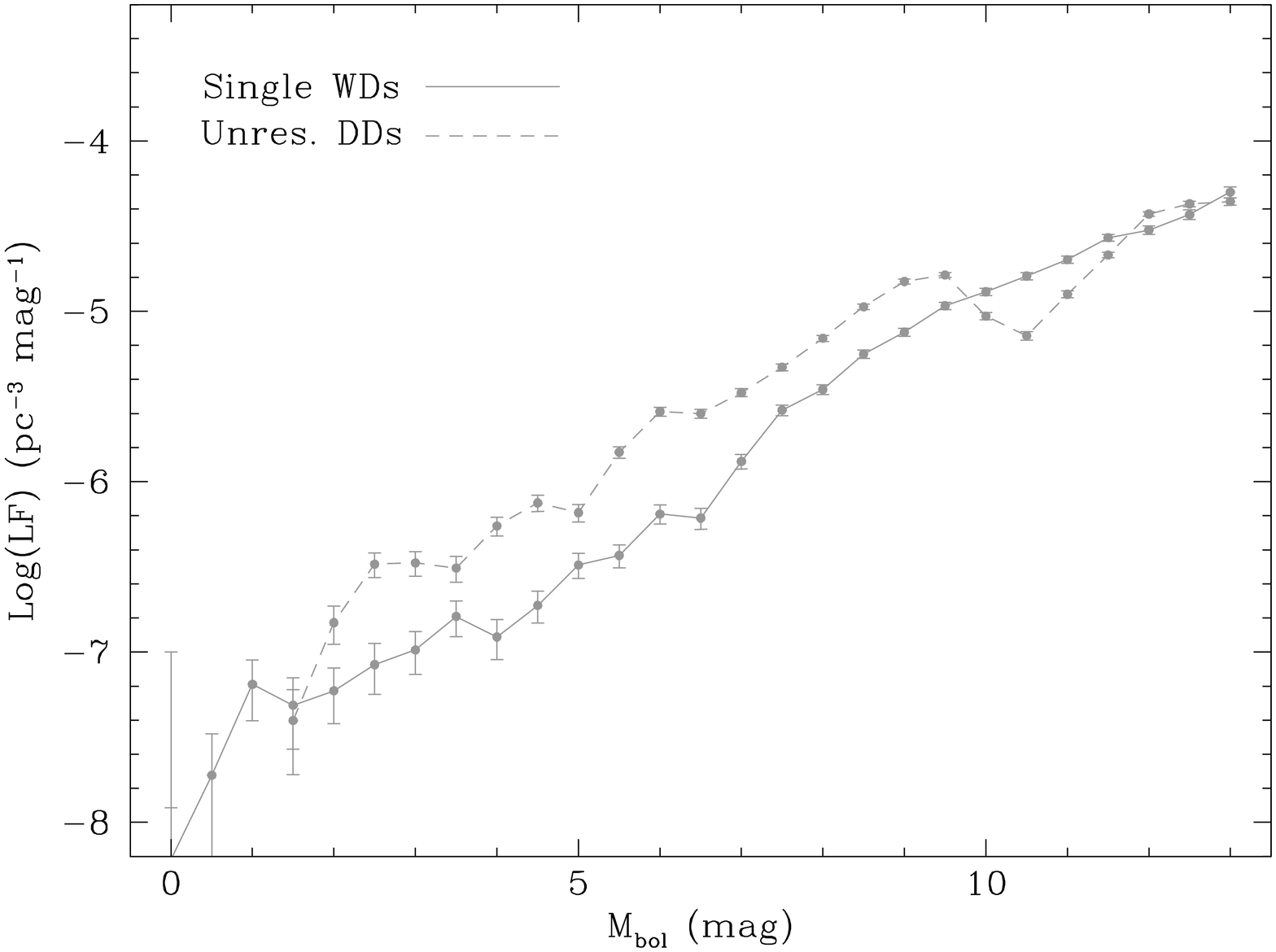}
\includegraphics[angle=-90,width=\columnwidth]{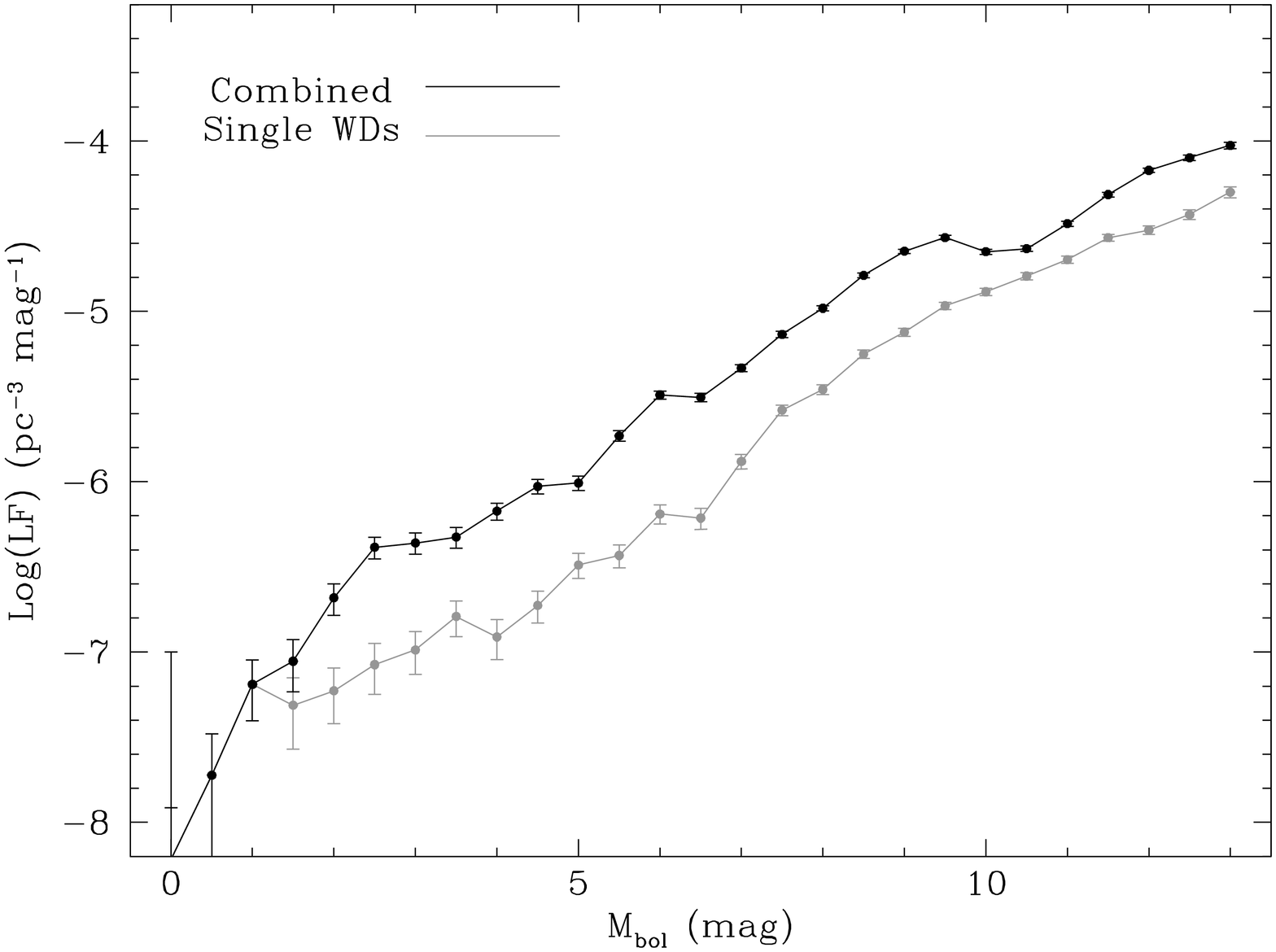}
\includegraphics[angle=-90,width=\columnwidth]{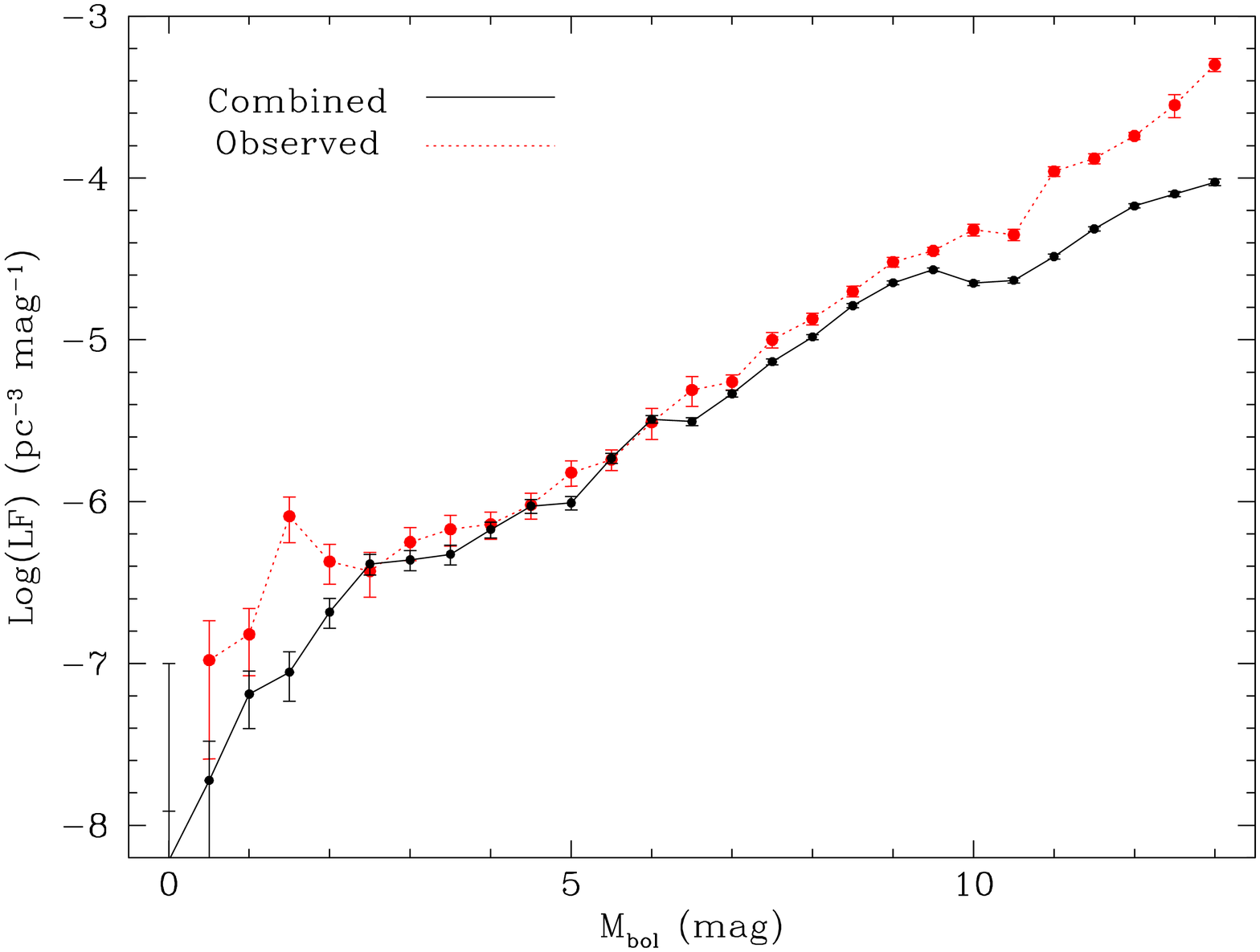}
    \caption{The same as Figure\,\ref{f-lfsdss1}  but for the the model
      that  produces   the  highest  fraction  of   unresolved  double
      degenerates.}
\label{f-lfsdss2}
\end{figure}

\begin{figure}
\includegraphics[angle=-90,width=\columnwidth]{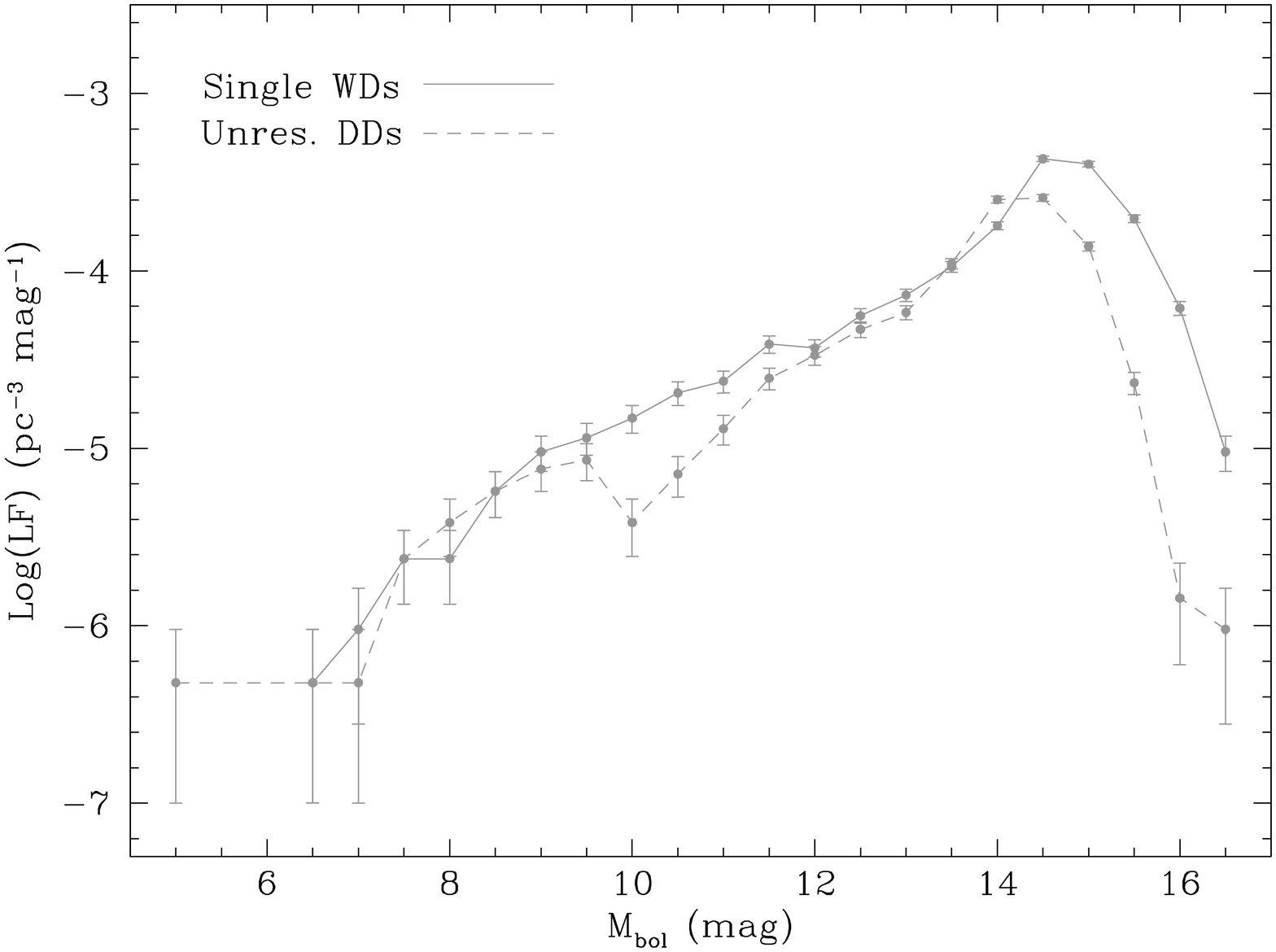}
\includegraphics[angle=-90,width=\columnwidth]{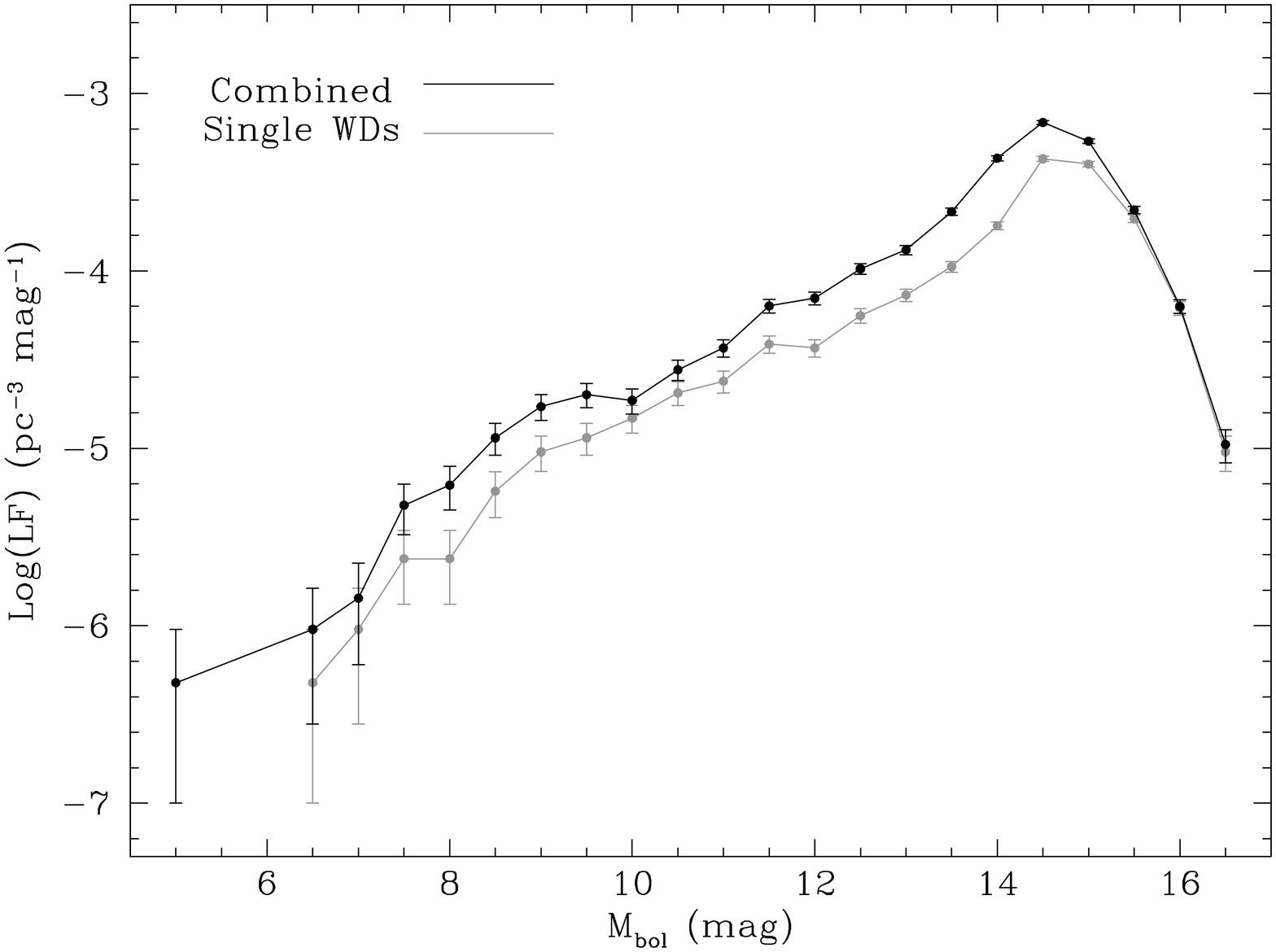}
\includegraphics[angle=-90,width=\columnwidth]{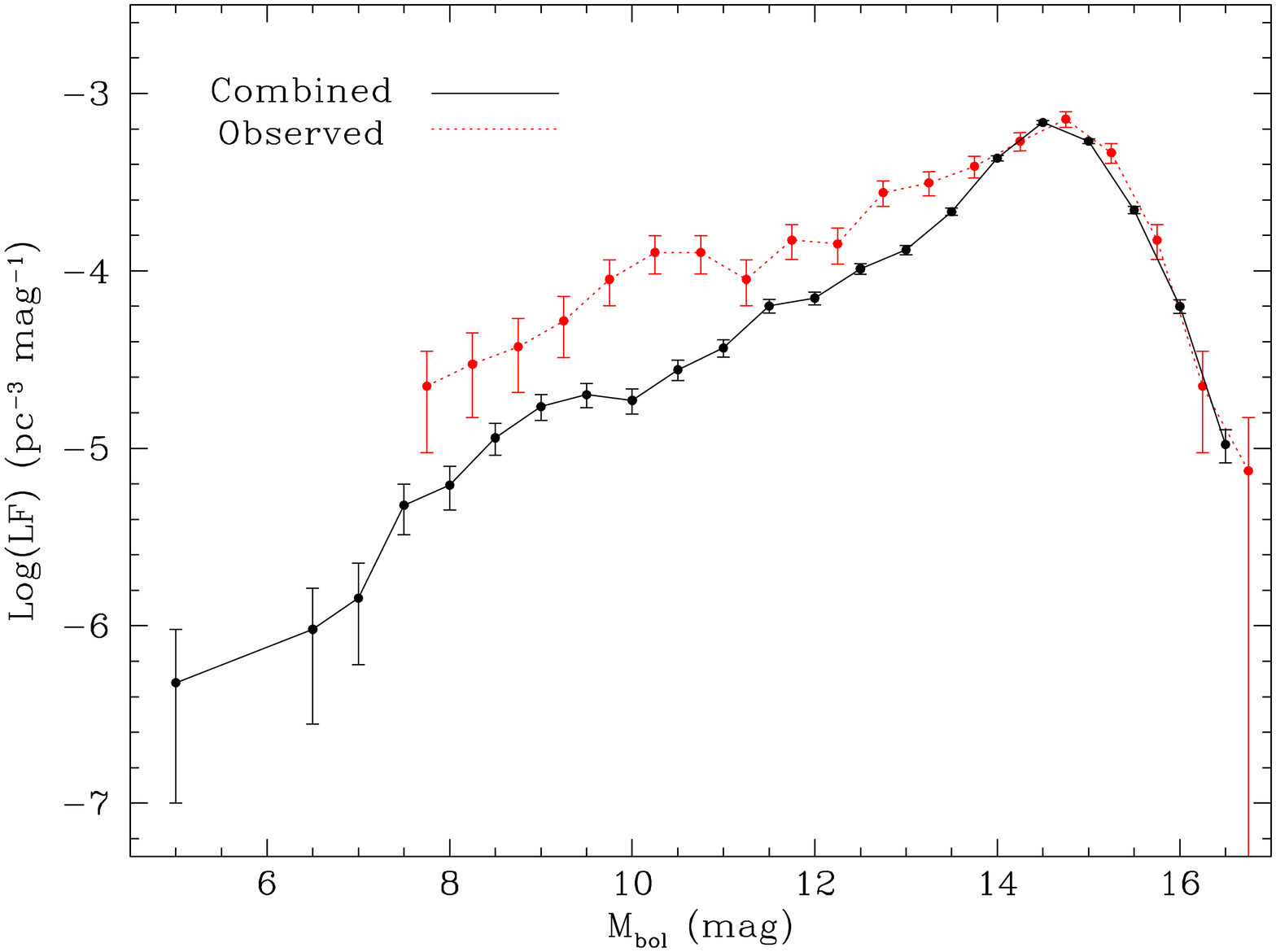}
    \caption{The same as Figure\,\ref{f-lfgaia1} but for the the model
      that  produces   the  highest  fraction  of   unresolved  double
      degenerates.}
\label{f-lfgaia2}
\end{figure}

We  performed a  new  simulation  to investigate  to  what extent  the
highest expected fraction of unresolved double WDs contaminates the WD
luminosity function. To  that end, we modified two  assumptions in our
standard model:  (1) the  binary fraction was  increased from  50\% to
85\% and (2) a mass ratio distribution of the form $n(q)\propto q$ was
adopted, where $q$  is the ratio of the secondary  and primary masses,
instead of a  uniform mass ratio distribution.   The two modifications
were based on a Monte Carlo simulator study that compares the fraction
of synthetic resolved WD binaries  to the observed fraction and allows
for deviations between the computed and  the observed values up to one
order of  magnitude (Canals et  al., in prep.).  The  first assumption
increased the number  of initial main sequence  binaries and therefore
the number  of resulting double  WDs. The  second implied that  it was
more  likely  to form  binaries  of  similar component  masses,  which
favoured the secondary  stars to be more massive and  hence to be able
to  evolve out  of  the main  sequence within  the  Hubble time,  thus
producing  a  larger  number  of  close double  WDs.   The  number  of
unresolved  double  degenerates  we  obtained  using  this  model  was
approximately  double that  of  our standard  model.   We derived  the
corresponding SDSS  and \emph{Gaia}  luminosity functions in  the same
way  as described  above for  our standard  model.  The  functions are
illustrated  in Figures\,\ref{f-lfsdss2}  and \ref{f-lfgaia2}  for the
SDSS  and \emph{Gaia}  sample, respectively.  The ratio  of the  space
densities of single WDs to unresolved double degenerates obtained with
this new model is illustrated in grey in Figure\,\ref{f-fraction}.

Because  of the  high binary  fraction assumed,  the space  density of
single SDSS  WDs was  lower than  the one  derived for  the unresolved
double degenerate binaries for most M$_\mathrm{bol}$ bins (see the top
panel     of    Figure\,\ref{f-lfsdss2}     and    top     panel    of
Figure\,\ref{f-fraction}).  As  a consequence,  the synthetic  SDSS WD
luminosity function is clearly affected by the inclusion of this extra
number   of   double   degenerates    (see   the   middle   panel   of
Figure\,\ref{f-lfsdss2}).  If  this was a  valid model, then  the star
formation rate  derived from  the SDSS  luminosity function  should be
taken  with  extreme  caution  due to  the  large  contamination  from
unresolved  double degenerates.   However, the  total synthetic  space
density  (single  WDs  plus  unresolved  double  degenerated)  clearly
underestimates  the space  density for  the observed  SDSS WDs  at the
bright  (M$_\mathrm{bol}<2$ mag)  and  faint (M$_\mathrm{bol}>9$  mag)
bins (see the bottom panel of Figure\,\ref{f-lfsdss2}).  This suggests
the high  number of  unresolved double  degenerates predicted  by this
model is unrealistic, most likely  because the adopted assumptions, as
compared  to   our  standard   model,  are   less  supported   by  the
observational data.

\begin{figure}
\includegraphics[angle=-90,width=\columnwidth]{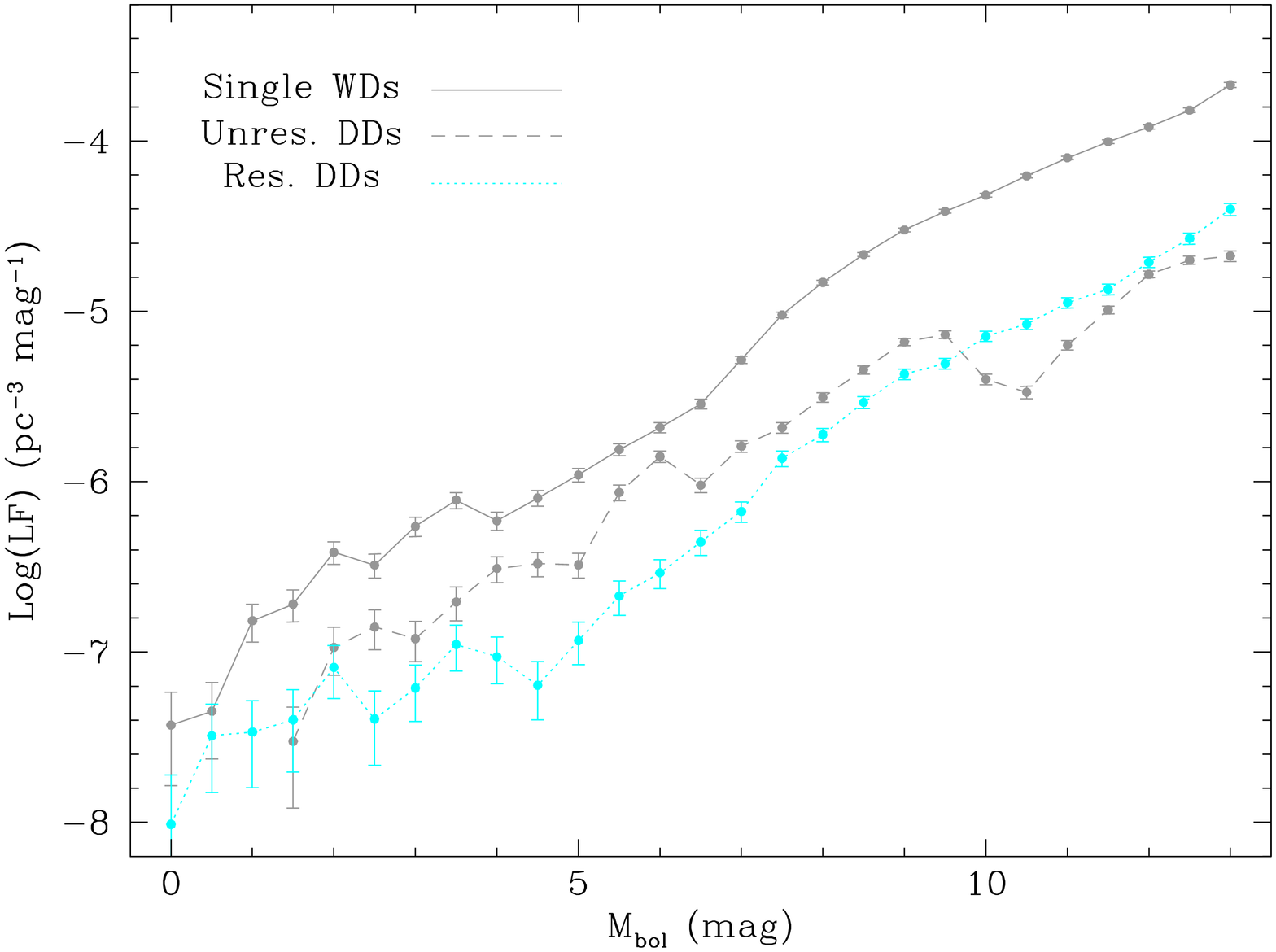}
\includegraphics[angle=-90,width=\columnwidth]{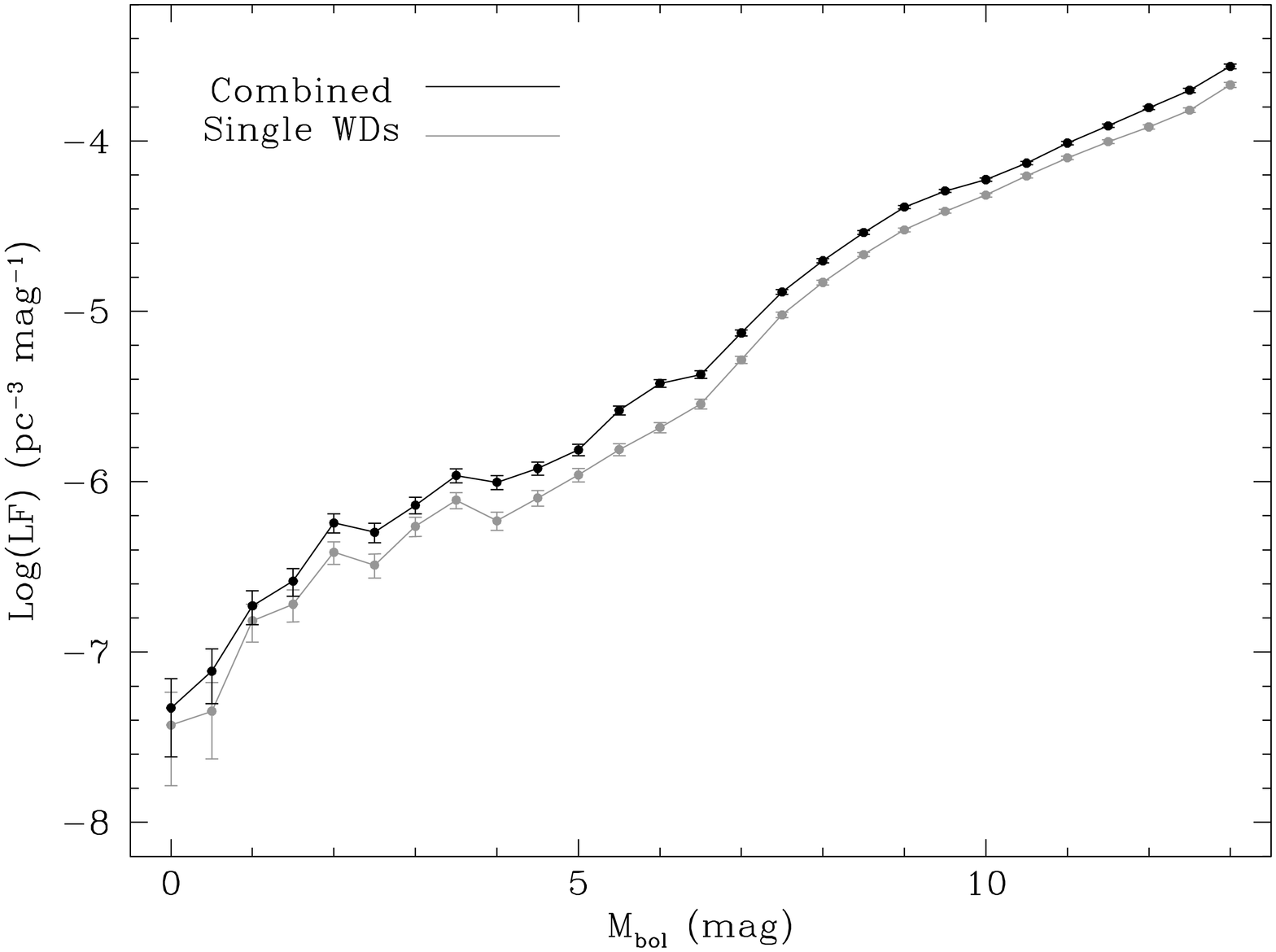}
\includegraphics[angle=-90,width=\columnwidth]{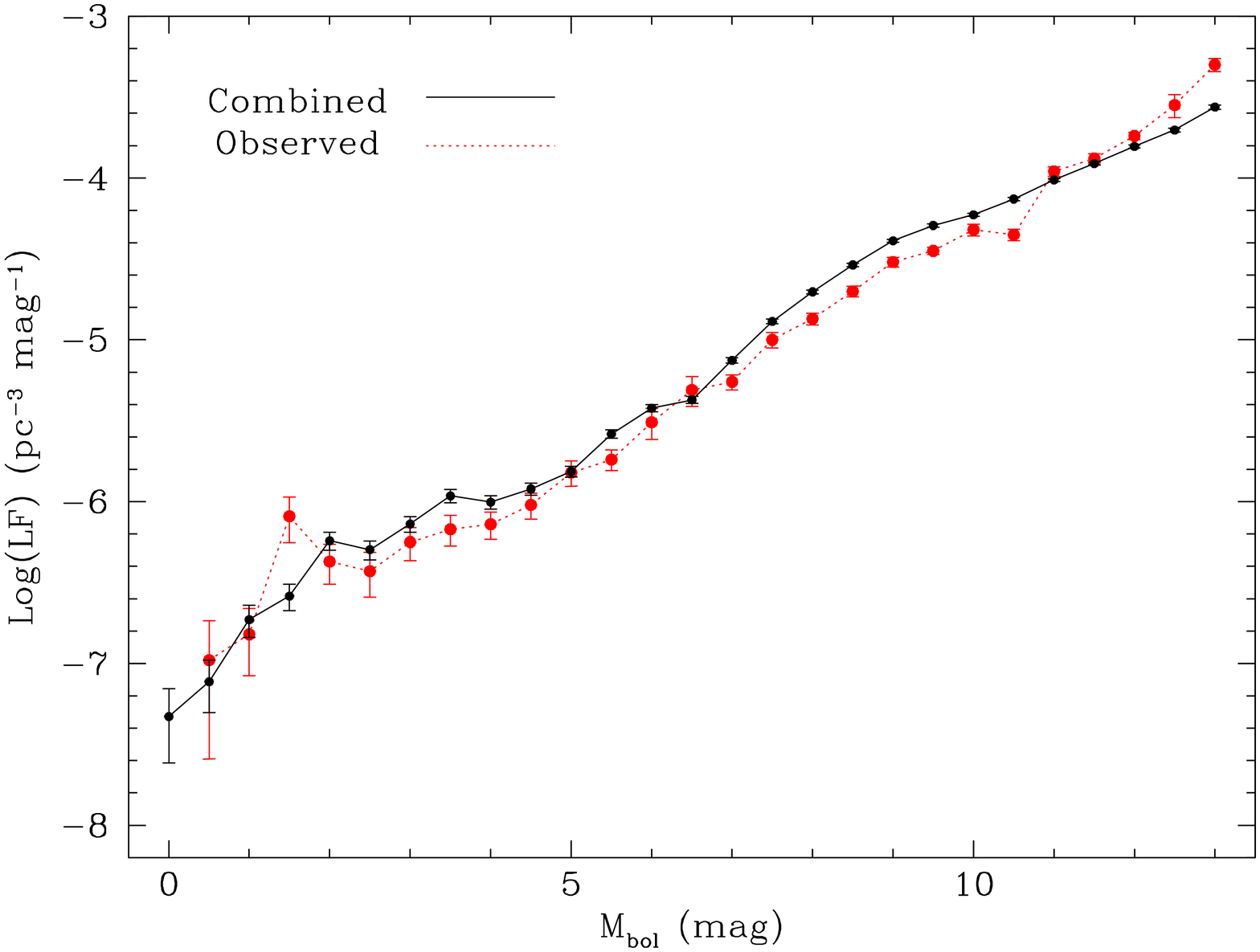}
\caption{The same  as Figure\,\ref{f-lfsdss1} but including  the space
  density contribution from  resolved double WDs (cyan  dotted line in
  the top-right panel).}
\label{f-lfsdss3}
\end{figure}

\begin{figure}
\includegraphics[angle=-90,width=\columnwidth]{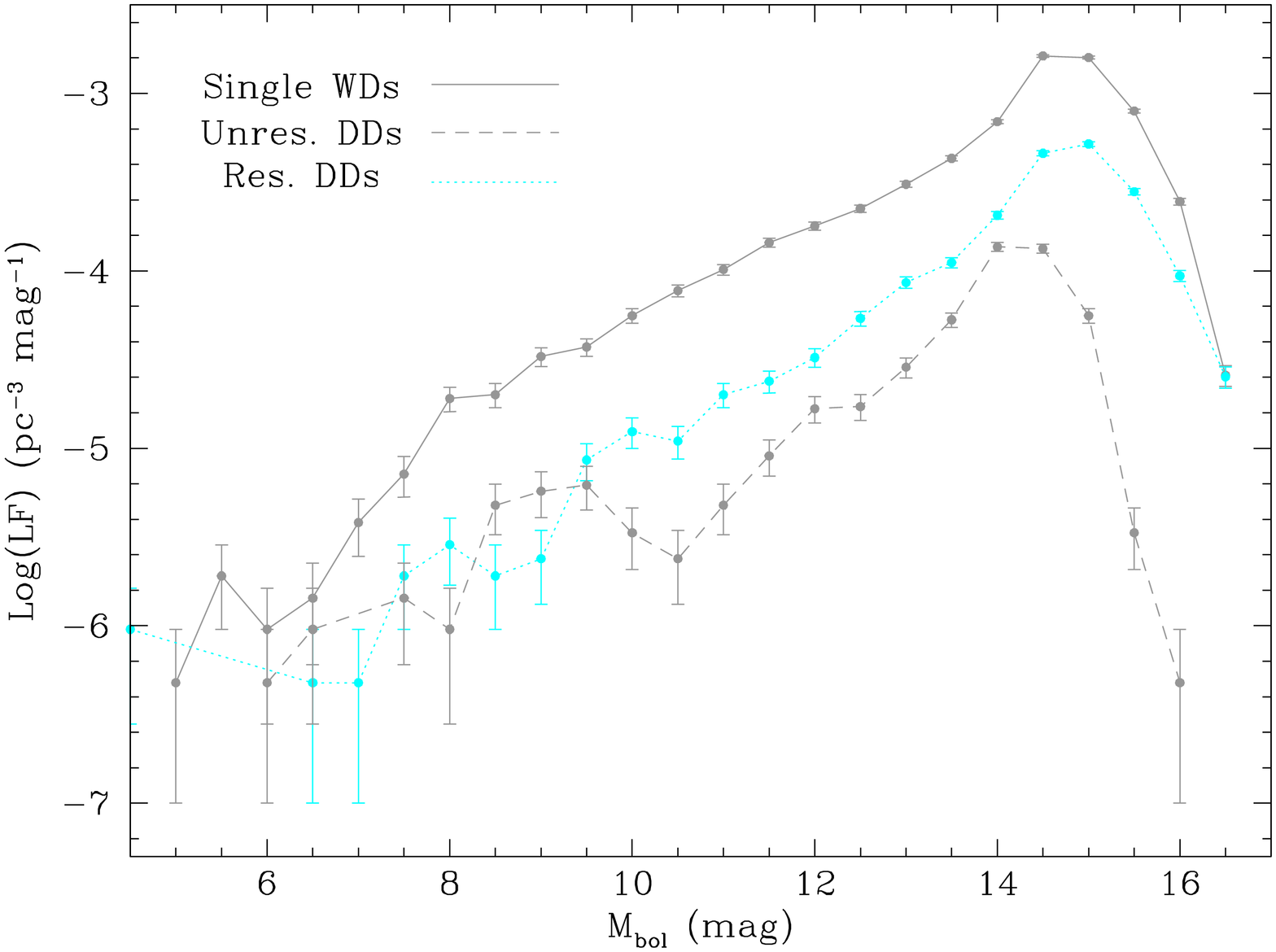}
\includegraphics[angle=-90,width=\columnwidth]{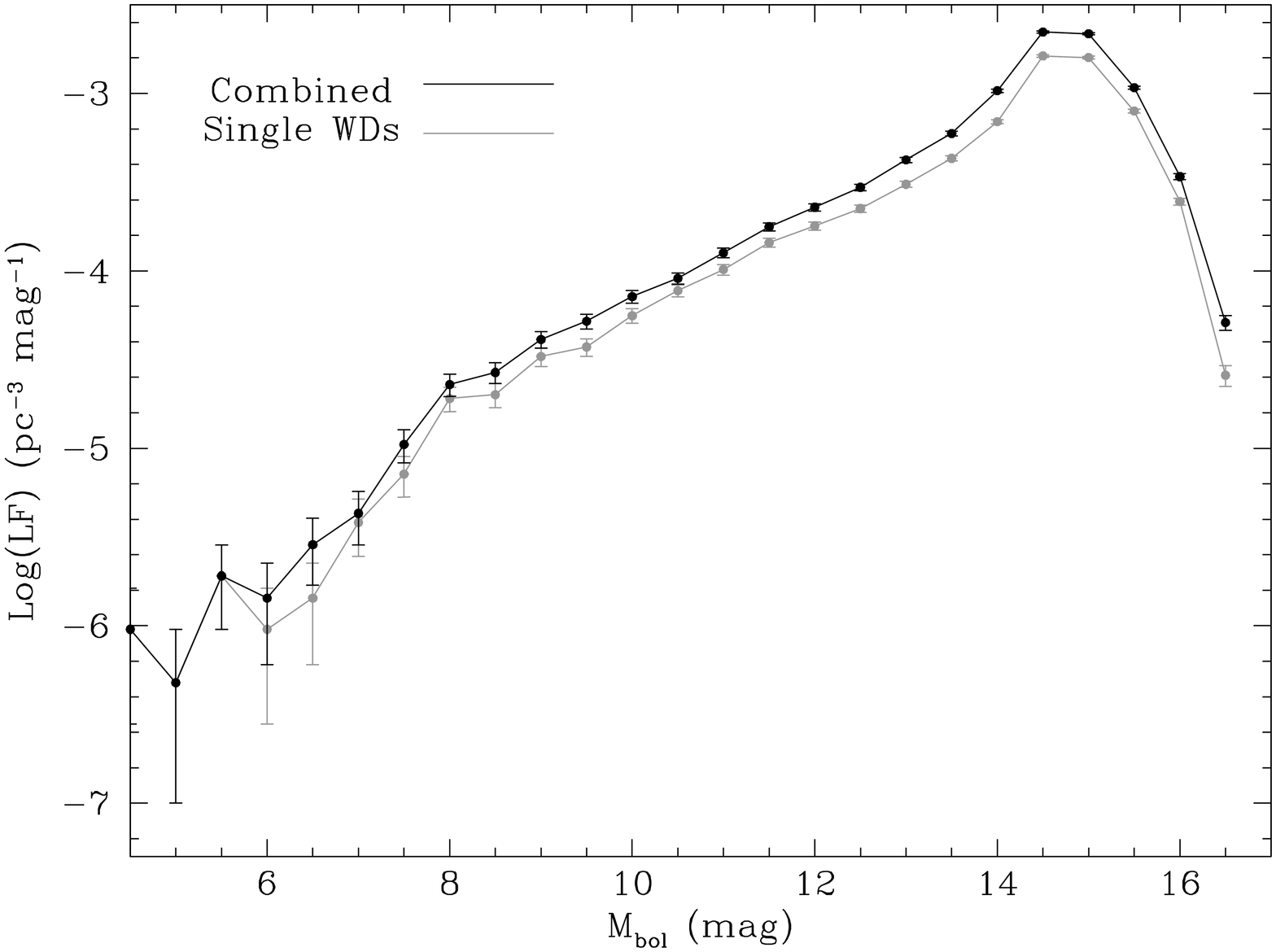}
\includegraphics[angle=-90,width=\columnwidth]{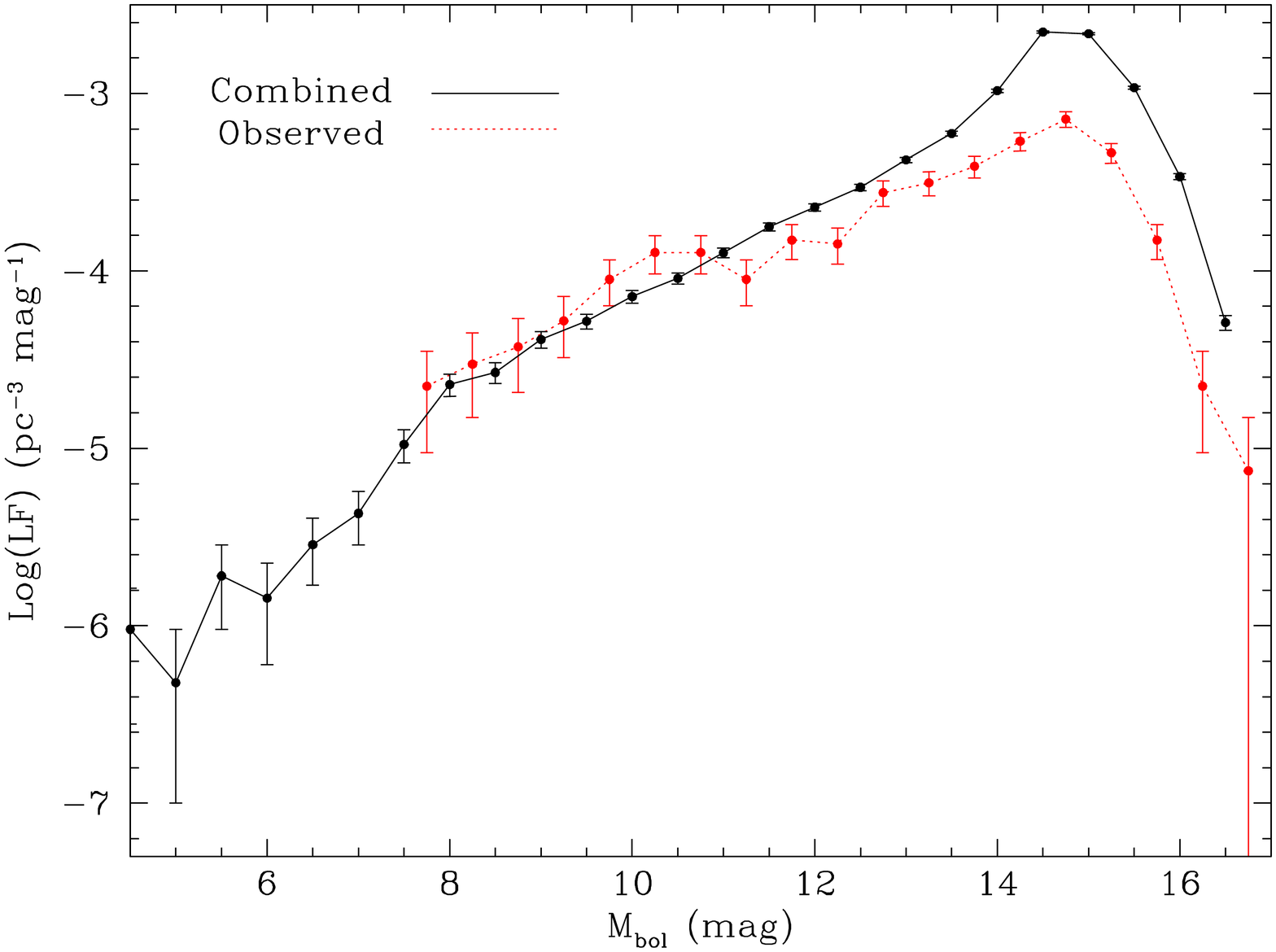}
\caption{The same  as Figure\,\ref{f-lfgaia1} but including  the space
  density contribution from  resolved double WDs (cyan  dotted line in
  the top-right panel).}
\label{f-lfgaia3}
\end{figure}

As  expected,  the  number   of  unresolved  double  degenerates  also
increased in the \emph{Gaia} synthetic  sample. However, in this case,
the  space density  of single  WDs was  higher or  similar to  the one
derived for unresolved WD binaries  in most M$_\mathrm{bol}$ bins (see
the  top   panel  of  Figure\,\ref{f-lfgaia2}  and   bottom  panel  of
Figure\,\ref{f-fraction}). The  effects of unresolved WD  binaries are
clearly  noticeable  at  all  bolometric  magnitude  bins  except  for
M$_\mathrm{bol}>15$     mag    (see     the     middle    panel     of
Figure\,\ref{f-lfgaia2}).    This  implies   that  even   the  highest
contamination   of   unresolved    double   degenerates   within   the
volume-limited  \emph{Gaia}  sample  is expected  to  have  negligible
effects on the  implied age of the Galactic disc  from the analysis of
the  cut-off  of the  WD  luminosity  function.  However,  this  model
implies   contamination   of   unresolved  double   degenerates   that
\emph{will}  affect the  derived star  formation history,  since these
unresolved binaries  would be considered  as single WDs.

We  conclude the  adopted model  in this  section is  most likely  not
realistic.  However, the  exercise performed here gives us  an idea of
the  maximum  contamination  one  can expect  from  unresolved  double
degenerates in both the SDSS and \emph{Gaia} WD luminosity functions.

\subsection{Resolved double degenerates}

We    followed     the    same     procedures    as     outlined    in
Section\,\ref{s-unresol} to  derive the effect of  resolved double WDs
in the  SDSS and \emph{Gaia}  luminosity functions. We  emphasise that
each binary component was considered  as a single WD which contributed
in space  density to the luminosity  functions.  The number of  WDs in
resolved double WDs  in our synthetic samples that  contributed to the
luminosity  functions  were  2,817  for  the  SDSS  sample\footnote{As
  explained  in Section\,\ref{s-unresol},  the SDSS  synthetic samples
  are filtered according to a colour  cut that defines the location of
  spectroscopic SDSS DA WDs.  This  implies that, on occasions, one WD
  in a resolved binary passes the filters but the other does not. This
  explains why  the resulting  number of  WDs is  an odd  number.} and
4,090 for the \emph{Gaia} sample.  As expected, the number of resolved
WDs in  the \emph{Gaia} sample  was larger  due to the  better angular
resolution (0.5" compared to 1" for SDSS) and smaller volume.  The top
panels   of   Figures\,\ref{f-lfsdss3}  (SDSS)   and   \ref{f-lfgaia3}
(\emph{Gaia}) show  the resolved double WD  luminosity functions (cyan
dotted lines) and  the middle panels display the  combined (single WDs
plus  unresolved  and  resolved   double  degenerates)  functions.   A
comparison  of  the combined  functions  with  the observed  ones  are
provided in the bottom panels of the same figures.

Although the space densities increase due to the inclusion of resolved
double  degenerates  in  both  samples,  the  shape  of  the  combined
synthetic  luminosity  functions  are  not  too  badly  affected  when
compared to those derived from our  standard model (see the middle and
bottom panels of  Figures\,\ref{f-lfsdss1} and \ref{f-lfgaia1}).  This
implies the impact of resolved double degenerates in the WD luminosity
function is  low. In any case,  it is important to  emphasise that the
binary components  in resolved  binaries can  be easily  identified if
parallaxes     and    proper     motions     are    available     (see
e.g. \citealt{Badry2018}  for the  \emph{Gaia} sample),  which implies
one can easily exclude these  binaries from the observational analysis
and thus avoid any possible contamination from such systems.

\subsection{White dwarf plus main sequence binaries}

In  the present  work we  have only  considered the  double WD  binary
population and  thus have excluded binaries  consisting of a WD  and a
main sequence  companion, i.e.   WDMS binaries, which  may contaminate
the WD luminosity function too.

Unresolved WDMS  binaries are generally  easy to identify  either from
colours and/or spectra.  Hence, these binaries will only contribute to
the WD luminosity function if  the MS companions are considerably less
luminous  than the  WDs. That  is, the  colours and  spectra of  these
systems  are very  similar to  those of  single WDs.   For the  20\,pc
sample,  \citet{Too17}  predicted,  using  the same  models  as  here,
$\simeq$4--8  unresolved  double  WDs   as  compared  to  $\simeq$1--2
unresolved WDMS binaries. For the  majority of these WDMS binaries the
flux was  completely dominated  by the  main sequence  companion, such
that we would  not confuse it for  being a single WD. This  is in line
with  the results  obtained here,  which indicate  that the  number of
unresolved WDMS  binaries in which  the WD clearly dominates  the flux
emission  is one  order of  magnitude  lower than  that of  unresolved
double degenerates.   Hence, unresolved WDMS binaries  clearly have no
effect in altering the synthetic WD luminosity functions.

For the resolved  WDMS binaries, the WDs can be  considered to be part
of the  same population as  that of single  WDs. By adding  those, the
only  noticeable effect  would be  to  increase the  space density  of
single WDs  and to thus lower  the fraction of double  degenerates per
bolometric  magnitude bin.   This would  would also  correspond to  an
increase in space density implied by the combined luminosity function.
However,  as   mentioned  above,  resolved  binaries   can  easily  be
identified  using  \emph{Gaia}  astrometric  data  and  can  hence  be
excluded from any observational analysis.

\section{Summary and conclusions}

In  this  work  we  have  studied  the  impact  of  unresolved  double
degenerates in  the white dwarf  luminosity function.  To that  end we
have  simulated  the single  and  the  double degenerate  white  dwarf
populations in  the solar  neighbourhood using a  population synthesis
approach.  The synthetic samples  have been obtained using assumptions
that best agree with the observational data, i.e.  our standard model.
Unresolved double white  dwarfs have been analysed in the  same was as
if   they  were   single   point  sources   of  fictitious   effective
temperatures,  surface  gravities   and  bolometric  magnitudes.   The
synthetic luminosity  functions have been obtained  by considering the
simulated   samples   as  if   they   were   observed  by   the   SDSS
magnitude-limited and \emph{Gaia} volume-limited surveys.

Under the assumptions  of our standard model we find  that the effects
of  unresolved  double  degenerates  in  the  white  dwarf  luminosity
function are nearly negligible for the \emph{Gaia} sample. Conversely,
for  the SDSS  sample  the contamination  of  unresolved double  white
dwarfs is significant at the  brighter bins of the luminosity function
(M$_\mathrm{bol}$<6.5 mag), reaching a  maximum of $\simeq$40\% of the
total white  dwarf population at M$_\mathrm{bol}=$6  mag. This implies
that  the age  of the  Galactic  disc as  well as  the star  formation
history that will eventually be  derived from the observed \emph{Gaia}
population  are  not expected  to  be  affected by  unresolved  double
degenerates.   However, the  star formation  history derived  from the
SDSS sample, particularly  during the most recent era,  is expected to
be  affected  to  some  extent  (up to  a  40\%  uncertainty)  by  the
contamination of these binaries.

We have considered an additional  model that maximizes the fraction of
unresolved double degenerates, with the  aim of evaluating the highest
impact  of  such  binaries  in  the  SDSS  and  \emph{Gaia}  synthetic
luminosity functions. Under the assumptions  of this model, the number
of double  white dwarfs is twice  that implied by our  standard model.
Cosequently,  the  shape  of  the luminosity  functions  are  severely
affected and  the star formation  histories derived from  both samples
(SDSS and \emph{Gaia}) would be  strongly affected. However, even with
the highest fraction of unresolved binaries considered, the age of the
Galactic disc derived  from the cut-off of  the \emph{Gaia} luminosity
function is not expected to be affected.  We also find that this model
seems to be  unrealistic, since the assumptions adopted  are in poorer
agreement with observations.

Finally, we  evaluated the effects  of resolved double  degenerates as
well  as white  dwarf plus  main sequence  binaries in  the luminosity
functions.  Resolved binaries  are found to slightly  modify the shape
of the luminosity functions. Nonetheless, these binaries can be easily
identified observationally  and thus  excluded from the  analysis.  In
the  case  of unresolved  white  dwarf-main  sequence binaries,  their
impact  on  the  white  dwarf  luminosity  function  is  found  to  be
negligible.

\section*{Acknowledgements}

The  authors are  greatly indebted  to D.  Koester for  developing the
white  dwarf   mode  atmosphere  spectra   used  in  this   work.  ARM
acknowledges  support  from  the  MINECO under  the  Ram\'on  y  Cajal
programme (RYC-2016-20254). ARM and S. Torres acknowledge support from
the AYA2017-86274-P grant and the  AGAUR grant SGR-661/2017. S. Toonen
acknowledges support from the  Netherlands Research Council NWO (grant
VENI [nr. 639.041.645]).










\bsp 
\label{lastpage}
\end{document}